\documentclass[useAMS]{cJAS2e}
\usepackage{natbib}
\RequirePackage[colorlinks,citecolor=blue,urlcolor=blue]{hyperref}
\hypersetup{breaklinks=true}  %% nicer for printable/submittable version

\usepackage{graphicx}
\usepackage{url}

\RequirePackage{hypernat}

\newcommand\tabfiginserted[1]{#1} \newcommand\tabfiglater[1]{} 

\newcommand\vijhat{\widetilde{v_{ij}}}
\newcommand\vikjhat{\widetilde{v_{{i_k}j}}}
\newcommand\viprimejhat{\widetilde{v_{i_{k'}j}}}
\newcommand\nboot{n_{\mathrm{b}}}  % bootstrap confidence interval
\newcommand\cboot{c_{\mathrm{b}}}  % bootstrap confidence interval
\newcommand\cbootstar{c_{\mathrm{b}}^*}  % bootstrap confidence interval
\newcommand\cempir{c_{\mathrm{e}}}  % empirical confidence interval
\newcommand\alphaboot{\alpha_{\mathrm{b}}}  % boostrap alpha 
\newcommand\alphaempir{\alpha_{\mathrm{e}}}  % empirical alpha
\newcommand\pseven{p_7} % chance of K = 7...  
\newcommand\pKsevenallodd{p_{\mathrm{odd}}^{\mathrm{K}7}} % p overall for all 4 candidates
\newcommand\psevenabcKS{p_{6{\mathrm{biggest}}}^{\mathrm{KS}}} % p using   K = 7abc as a selector
\newcommand\psevenabcG{p_{6{\mathrm{biggest}}}^{\mathrm{G}}} % p using   K = 7abc as a selector
\newcommand\psevenabczeros{p_{d_2=d_3=d_4}} % p using  K = 7abc as a selector
\newcommand\psevena{p_{7d_2}} % p using  K = 7a as a selector
\newcommand\pall{p_{\mathrm{all}}} % p overall 
\newcommand\gtapprox{\,\lower.6ex\hbox{$\buildrel >\over \sim$} \, }
\newcommand\ltapprox{\,\lower.6ex\hbox{$\buildrel <\over \sim$} \, }

\newcommand\notea{\ifmmode{}^{\mathrm a}\else${}^{\mathrm a}$\fi}
\newcommand\noteb{\ifmmode{}^{\mathrm b}\else${}^{\mathrm b}$\fi}
\newcommand\notec{\ifmmode{}^{\mathrm c}\else${}^{\mathrm c}$\fi}
\newcommand\noted{\ifmmode{}^{\mathrm d}\else${}^{\mathrm d}$\fi}
\newcommand\notee{\ifmmode{}^{\mathrm e}\else${}^{\mathrm e}$\fi}
\newcommand\notef{\ifmmode{}^{\mathrm f}\else${}^{\mathrm f}$\fi}
\newcommand\noteg{\ifmmode{}^{\mathrm g}\else${}^{\mathrm g}$\fi}
\newcommand\noteh{\ifmmode{}^{\mathrm h}\else${}^{\mathrm h}$\fi}
\newcommand\notei{\ifmmode{}^{\mathrm i}\else${}^{\mathrm i}$\fi}
\newcommand\notej{\ifmmode{}^{\mathrm j}\else${}^{\mathrm j}$\fi}

\newcommand\mybiblioauthorstyle[1]{#1}
\newcommand\mybibliotitlestyle[1]{{\em #1}}

\newcommand\postrefereechanges[1]{#1}    

\newcommand\mygraphwidthsquare{55mm} \newcommand\mygraphwidth{70mm}

\newcommand\fsigabs{
\begin{figure}
\centering 
\includegraphics[width=\mygraphwidth]{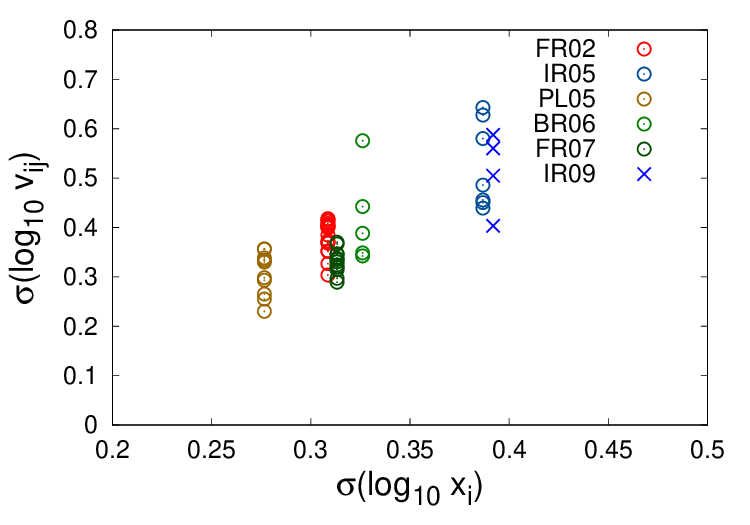}
\caption[]{Logarithmic widths 
(standard deviations)
of per-candidate vote distributions 
$v_{ij}$ vs widths of total vote distributions $x_i$. 
From left to right: 
Poland 2005,
France 2002,
France 2007,
Brazil 2006,
Iran 2005 ($\odot$);
Iran 2009 ($\times$). 
}
\label{f-sigabs}
\end{figure} 
} % of \newcommand\fsigabs

\newcommand\fsigrel{
\begin{figure}
\centering 
\includegraphics[width=\mygraphwidth]{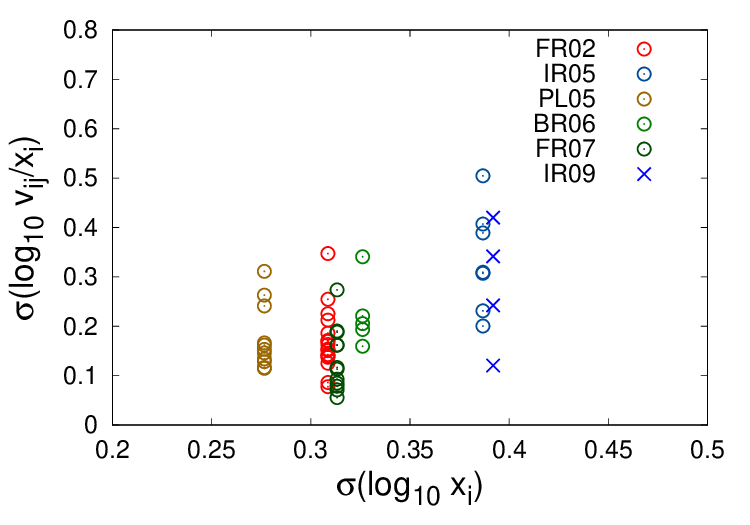}
\caption[]{As for 
Fig.~\protect\ref{f-sigabs}, 
logarithmic widths of per-candidate voting rates
$\widetilde{v_{ij}}$ vs widths of total vote distributions $x_i$. 
}
\label{f-sigrel}
\end{figure} 
} % of \newcommand\fsigrel

\newcommand\fbooteg{
\begin{figure}
\centering 
\includegraphics[width=\mygraphwidth]{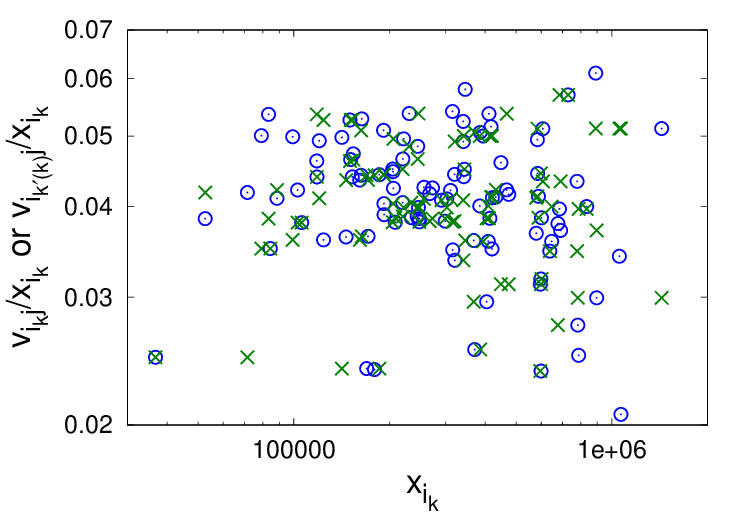}
\caption[]{Example of a local bootstrap realisation for 
candidate 1 in the French 2007 presidential-election first round, showing
official voting rates $\vijhat$ ($\odot$) and simulated 
voting rates $\viprimejhat$ ($\times$).
}
\label{f-booteg}
\end{figure} 
} % of \newcommand\fbooteg

\newcommand\fcalib{
\begin{figure}
\centering 
\includegraphics[width=\mygraphwidthsquare,bb=100 50 360 302]{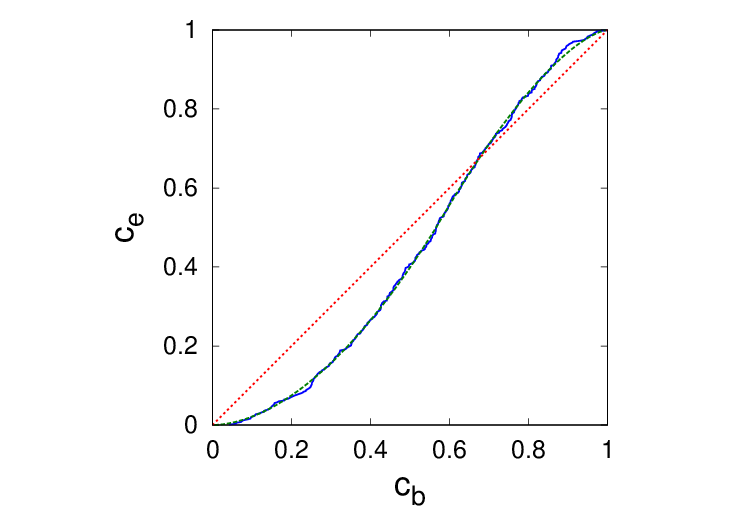}
\caption[]{Calibration of confidence levels using
the control data set.
The empirical confidence levels $\cempir$ are shown against
local bootstrap confidence levels $\cboot$ as a thick curve,
and the smooth fit is given by 
Eqs~(\protect\ref{e-cempir-cboot}) and 
(\protect\ref{e-calib-solution}). If the bootstrap confidence intervals
were unbiased and symmetric, then the relation $\cempir = \cboot$ 
would hold, as illustrated for comparison.
}
\label{f-calib}
\end{figure} 
} % of \newcommand\fcalib

\newcommand\fAconflev{
\begin{figure}
\centering 
\includegraphics[width=\mygraphwidth]{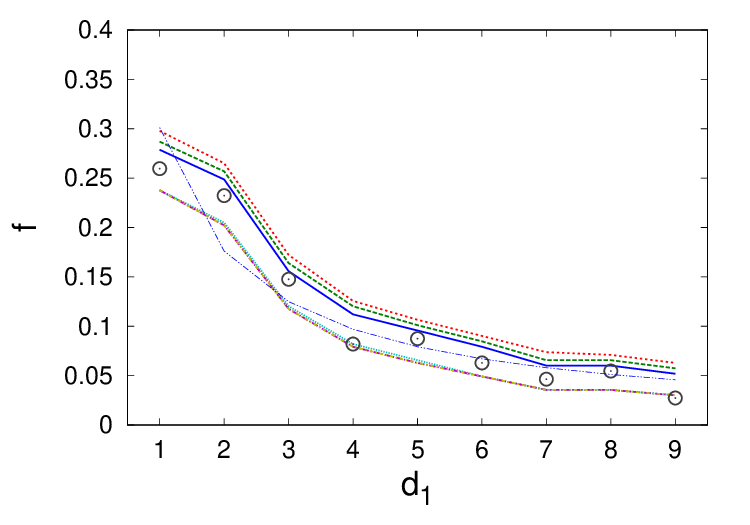}
\caption[]{Frequency 
distribution $f(d_1)$ of the first digit $d_1$ for 
candidate A 
($\odot$) in the Iranian 2009 presidential-election first round 
\protect\citep{MOI09}. Lower and upper confidence levels of
$\cempir = 0.05\%, 0.5\%, 2.5\%,
97.5\%, 99.5\%, 99.95\%$,
using the local bootstrap simulations
and the empirical calibration given in 
Eqs~(\protect\ref{e-cempir-cboot}) and 
(\protect\ref{e-calib-solution}), are shown.
The lower three confidence levels are close to one another and difficult
to distinguish in the plot.
The limiting case of the Benford's Law first-digit distribution
is shown as a thin line. 
}
\label{f-Aconflev}
\end{figure} 
} % of \newcommand\fAconflev

\newcommand\fRconflev{
\begin{figure}
\centering 
\includegraphics[width=\mygraphwidth]{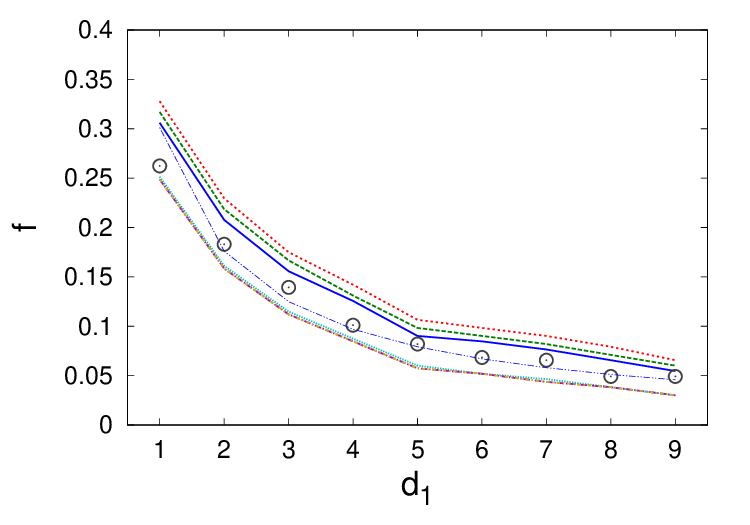}
\caption[]{First-digit frequency 
distribution $f(d_1)$ for candidate R, 
as for Fig.~\protect\ref{f-Aconflev}. 
}
\label{f-Rconflev}
\end{figure} 
} % of \newcommand\fRconflev

\newcommand\fKconflev{
\begin{figure}
\centering 
\includegraphics[width=\mygraphwidth]{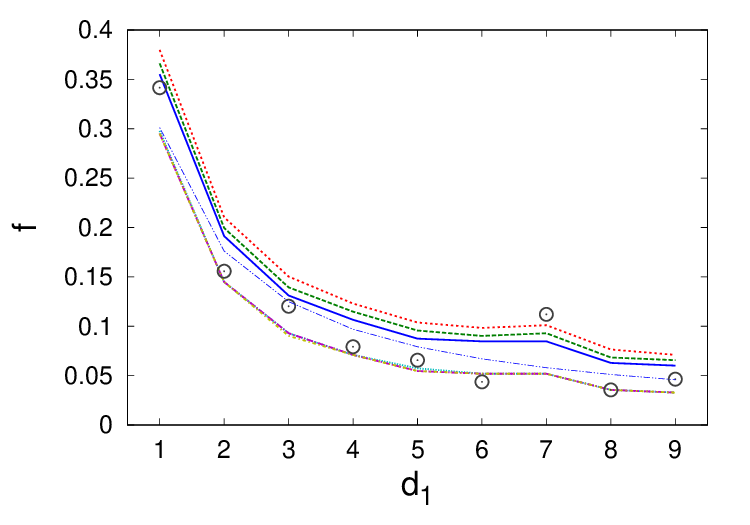}
\caption[]{First-digit frequency 
distribution $f(d_1)$ for candidate K, 
as for Fig.~\protect\ref{f-Aconflev}. 
}
\label{f-Kconflev}
\end{figure} 
} % of \newcommand\fKconflev

\newcommand\fMconflev{
\begin{figure}
\centering 
\includegraphics[width=\mygraphwidth]{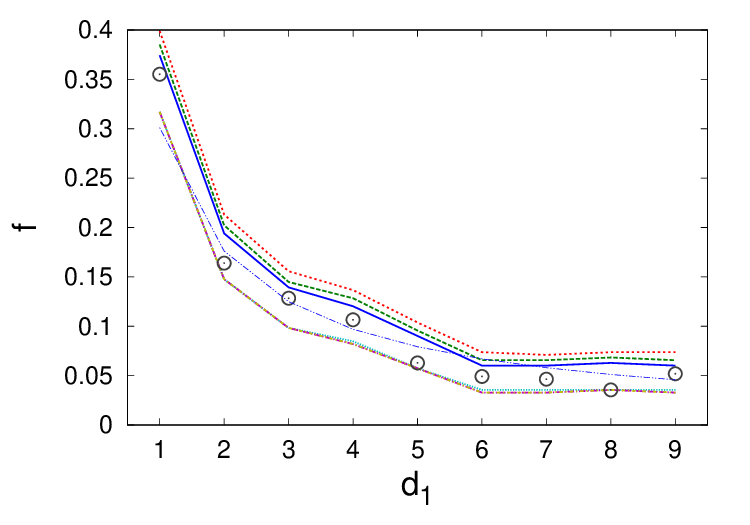}
\caption[]{First-digit frequency 
distribution $f(d_1)$ for candidate M, 
as for Fig.~\protect\ref{f-Aconflev}. 
}
\label{f-Mconflev}
\end{figure} 
} % of \newcommand\fMconflev

\newcommand\fprepolls{
\begin{figure}
\centering 
\includegraphics[width=\mygraphwidth]{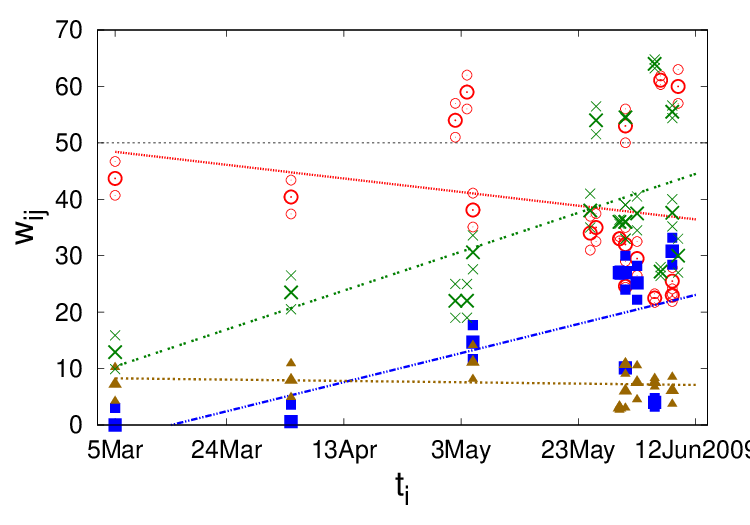}
\includegraphics[width=\mygraphwidth]{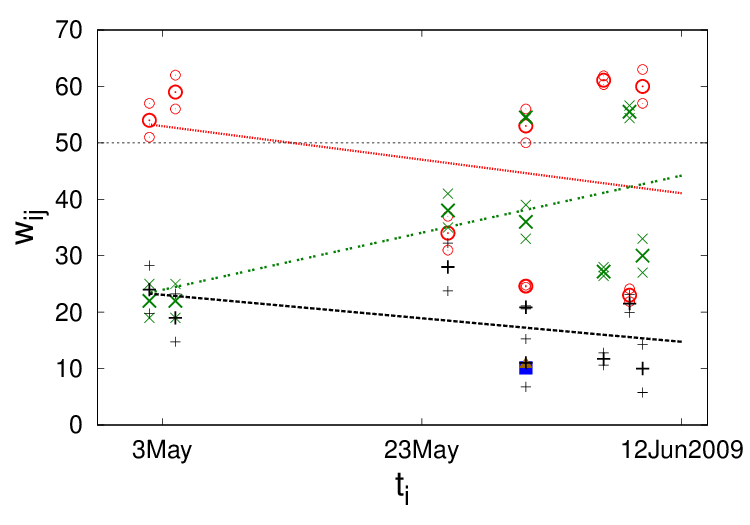}
\caption[]{Pre-election poll voting intentions $w_{ij}$\%
  (Table~\protect\ref{t-prepolls}) for 
  candidates $j=$ A ($\odot$),
  R (squares), K (triangles), M ($\times$), versus date
  $t_i$ in days, where  $100=$ 12 June 2009
  (the election date).  Error bars defined as $100\% /
  \sqrt{q_i}$ are shown as corresponding small/thin symbols, where $q_i :=
  1000$ if no value is stated in the table.  Linear least-squares
  equal-weighted fits to $w_{ij}(t_i)$ for each $j$-th candidate
  are shown, including extrapolations to 12 June. The 50\%
  threshold for winning the election is shown. 
  {\em Upper panel:} all four candidates, all polls; 
  {\em lower panel:} candidates A, M and RKO (total support for
  candidates R, K and ``other'' inferred from support for A and M)
  ($+$), excluding partially internet-based and publicly unarchived
  polls.
}
\label{f-prepolls}
\end{figure} 
} % of \newcommand\fprepolls

\newcommand\fhistKfolded{
\begin{figure}
\centering 
\includegraphics[width=\mygraphwidth]{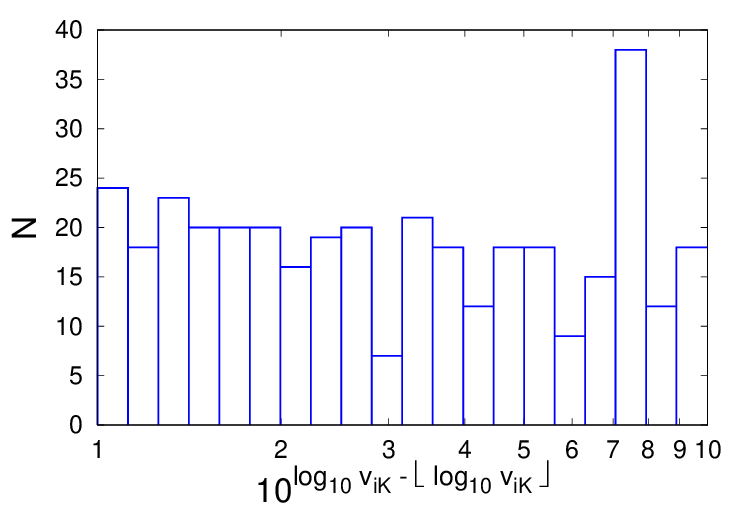}
\caption[]{Distribution of the folded logarithmic vote counts for 
candidate K (cf. Fig.~\protect\ref{f-Aconflev}, bottom-left).
}
\label{f-histKfolded}
\end{figure} 
} % of \newcommand\fhistKfolded

\newcommand\fhistall{
\begin{figure}
\centering 
\includegraphics[width=\mygraphwidth]{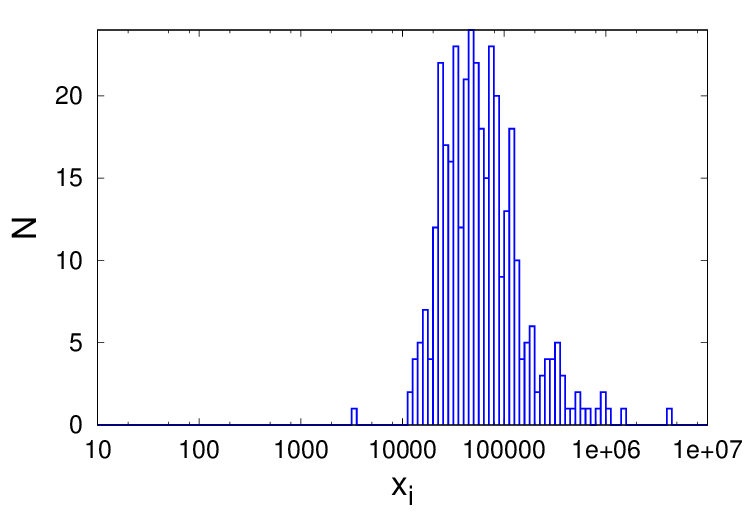}
\caption[]{Logarithmic distribution of the total vote counts 
$x_i$.
}
\label{f-histall}
\end{figure} 
} % of \newcommand\fhistall

\newcommand\fhistA{
\begin{figure}
\centering 
\includegraphics[width=\mygraphwidth]{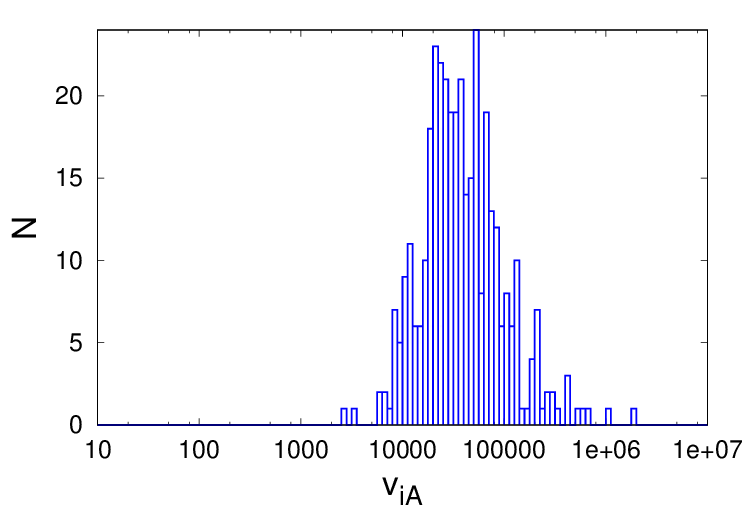}
\caption[]{Logarithmic distribution of 
the vote counts $v_{i\mathrm A}$
for candidate A.}
\label{f-histA}
\end{figure} 
} % of \newcommand\fhistA

\newcommand\fhistR{
\begin{figure}
\centering 
\includegraphics[width=\mygraphwidth]{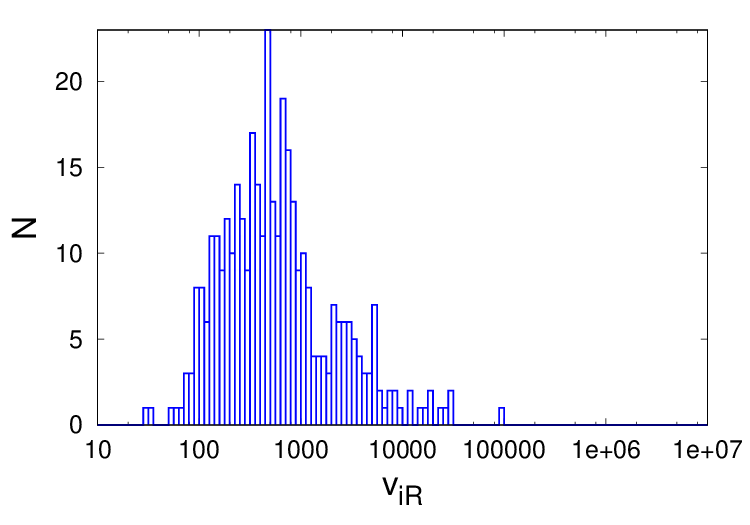}
\caption[]{Logarithmic distribution of $v_{i\mathrm R}$, the 
vote counts for R.
}
\label{f-histR}
\end{figure} 
} % of \newcommand\fhistR

\newcommand\fhistK{
\begin{figure}
\centering 
\includegraphics[width=\mygraphwidth]{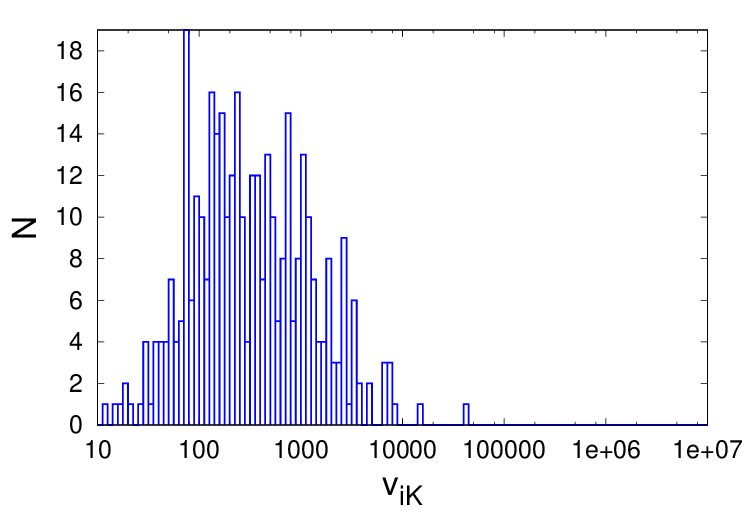}
\caption[]{Logarithmic distribution of $v_{i\mathrm K}$, the 
vote counts for K.  
}
\label{f-histK}
\end{figure} 
} % of \newcommand\fhistK

\newcommand\fhistM{
\begin{figure}
\centering 
\includegraphics[width=\mygraphwidth]{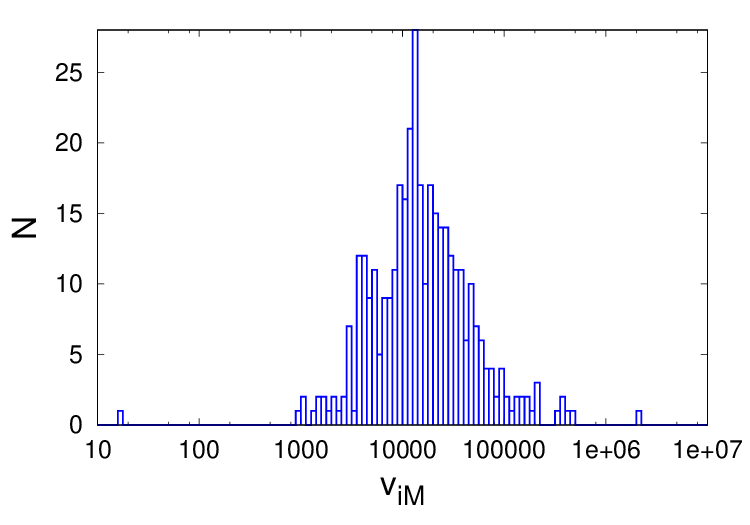}
\caption[]{Logarithmic distribution of $v_{i\mathrm M}$,
the 
vote counts for M.  
}
\label{f-histM}
\end{figure} 
} % of \newcommand\fhistM

\newcommand\fKsevenA{
\begin{figure}
\centering 
\includegraphics[width=\mygraphwidth]{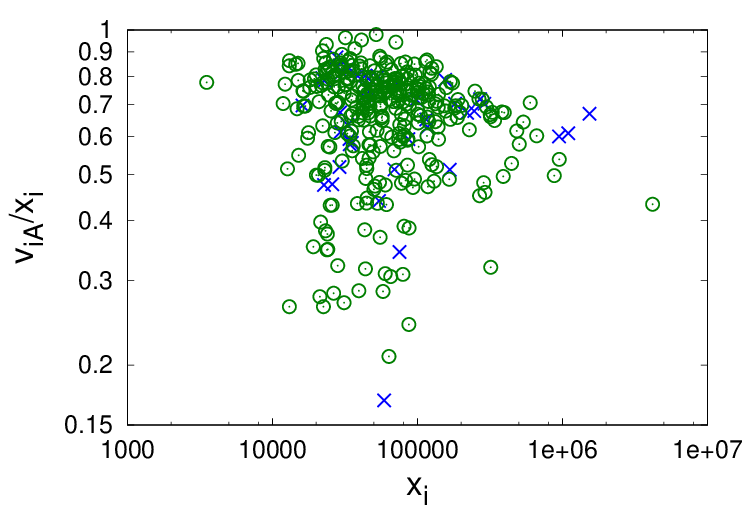}
\caption[]{Proportion of vote counts 
$v_{i\mathrm A}/x_i$ against total votes 
$v_j$, for candidate A. Voting areas are selected by whether 
K's vote count has the first digit 7 ($\times$) or another
digit ($\odot$). 
\protect\postrefereechanges{The vertical (logarithmic) scales
  differ greatly between this figure and 
  Figs~\protect\ref{f-KsevenR}--\protect\ref{f-KsevenM}, in order
  to show the structures of the distributions. To compare the
  widths of the $\log_{10} v_{ij}/x_i$ 
  distributions in the four figures, 
  see Fig.~\protect\ref{f-sigrel}.}
}
\label{f-KsevenA}
\end{figure} 
} % of \newcommand\fKsevenA

\newcommand\fKsevenR{
\begin{figure}
\centering 
\includegraphics[width=\mygraphwidth]{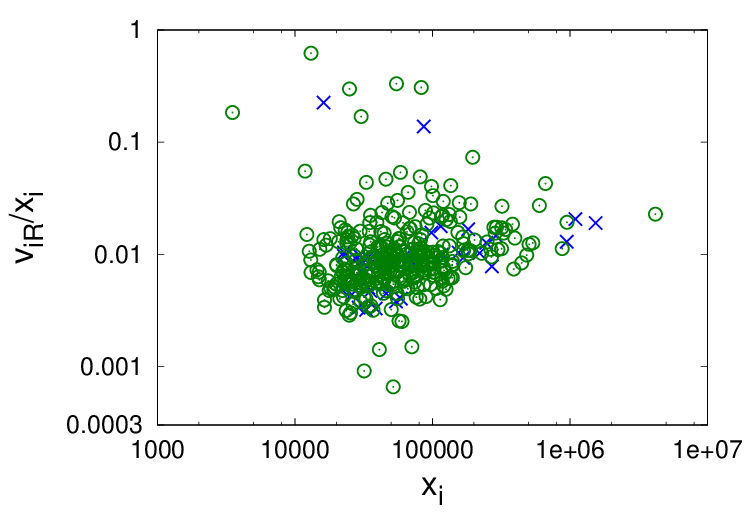}
\caption[]{R's proportions of vote counts, 
as for Fig.~\protect\ref{f-KsevenA}.
}
\label{f-KsevenR}
\end{figure} 
} % of \newcommand\fKsevenR

\newcommand\fKsevenK{
\begin{figure}
\centering 
\includegraphics[width=\mygraphwidth]{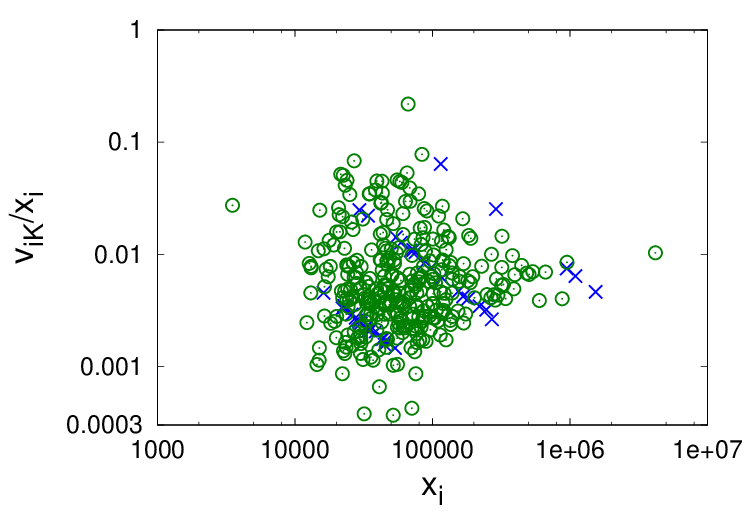}
\caption[]{K's proportions of vote counts, 
as for Fig.~\protect\ref{f-KsevenA}. The selection by
the first digit 7 is clear.
}
\label{f-KsevenK}
\end{figure} 
} % of \newcommand\fKsevenK

\newcommand\fKsevenM{
\begin{figure}
\centering 
\includegraphics[width=\mygraphwidth]{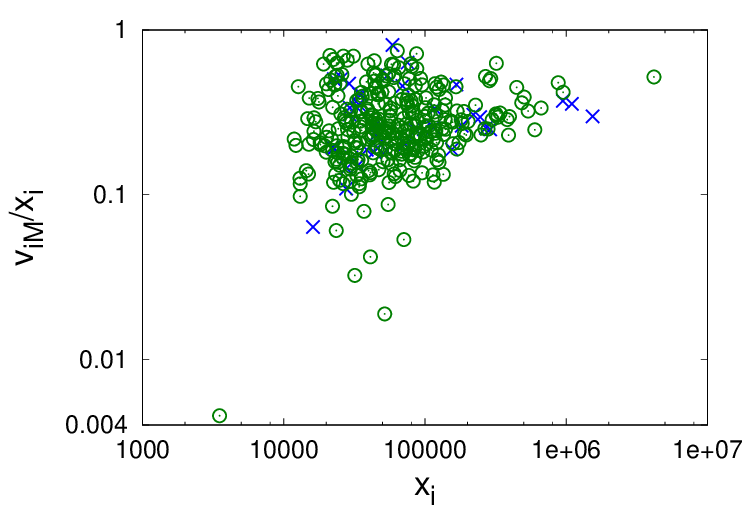}
\caption[]{M's proportions of vote counts, 
as for Fig.~\protect\ref{f-KsevenA}.
}
\label{f-KsevenM}
\end{figure} 
} % of \newcommand\fKsevenM

\newcommand\tpreselections{
\begin{table*}
  \caption{Recent presidential-election first rounds, each with
    $n$ candidates from $m$ voting areas, and a total of
    $\sum x_i$ votes. 
      \label{t-pres-elections}}
    $$\begin{array}{l c r r r l l} \hline 
      \mbox{country} & \mbox{date} & 
      \protect\postrefereechanges{m} & 
      \protect\postrefereechanges{n (n_0)}^{\mathrm{a}} &
      {\sum x_i }^{\mathrm{b}} 
      & \mbox{admin. division} & \mbox{source}\\
\hline 
      \mbox{France} & \mbox{2002-04-21} & 100 &  16 & 29,183,176 & \mbox{department} & \mbox{\protect\cite{MOIF02}} \\
      \mbox{Iran}  & \mbox{2005-06-17} & 326 & 7 & 28,155,678 & \mbox{shahrestan} & \mbox{\cite{Meb09b}} \\
      \mbox{Poland}  & \mbox{2005-10-09} & 379  & 11 (12) & 14,993,138
      & \mbox{powiat} & \mbox{\cite{PKW05}} \\
      \mbox{Brazil}  & \mbox{2006-10-01} & 2832 & 5(7)  & 94,785,276 &  \mbox{``electoral zone''}^{\mathrm c} & \mbox{\cite{TSEBr06}} \\
      \mbox{France}  & \mbox{2007-04-22} & 100 & 12 & 36,674,996 & \mbox{department} & \protect\mbox{\cite{MOIF07}} \\
      \hline
      \mbox{Iran}  & \mbox{2009-06-12} & 366 & 4 & 39,245,991 & \mbox{shahrestan} & \mbox{\cite{MOI09}} \\
\hline 
\end{array}$$
\tabnote{${}^{\mathrm{a}}$The analysis is carried out for 
\protect\postrefereechanges{$n$ candidates, where those
candidates among the original $n_0$} who have
 1\% or more votes of 1 or 0 are excluded 
(Section~\protect\ref{s-meth-calibration}).}
\tabnote{${}^{\mathrm{b}}$Total votes $\sum x_i$ 
are the sums of total votes per $i$-th region, $x_i$, stated in the source,
except for the Iranian 2005 election,
where $x_i$ are summed from the candidates' votes in the voting areas.
The total votes $\sum x_i$ may differ from the full total
due to groups not included in the main geographical regions, e.g. 
voters living abroad, on ships, etc.}
\tabnote{${}^{\mathrm{c}}$The ``electoral zones'' in the 
Superior Electoral Court's data are smaller than states and larger than
municipalities.}
\end{table*}
} %\newcommand\tpreselections

\newcommand\tprepolls{
\begin{table*}
  \caption{Intentions of $q_i$ voters 
    to vote for candidate $j$ with probability $w_{ij}$
    in the 2009 Iranian
    presidential first-round election, 
    in the $i$-th of 17 
    pre-election opinion polls \protect\citep{Gerash772009a}.
  \label{t-prepolls}}
    \begin{tabular}{l r  l l l l   l l } \hline 
      \mbox{date}{\notea}
      & $q_i$
      & $w_{i{\mathrm A}}$ & $w_{i{\mathrm R}}$ & $w_{i{\mathrm K}}$ & $w_{i{\mathrm M}}$
      & {source} & organ.{\noteb}
      \\
\hline
\rule{0ex}{2.5ex}   % strut 
  $<$  10 June 2009   &  -   &  60\%{\notec}  &  -  &  -  &  30\%{\notec}   &  \protect\cite{AlefPolls0318}    &  - \\
  $<$  9 June 2009    &   7900    &  23 \%   &  -  &  -  &  55.5\%{\noted}   &  \protect\cite{RoozonlinePolls0319}   &  - \\
  $<$  9 June 2009    &  1743    &  25.5 \%   &  30.8 \%  &  6.1 \%    &  37.6 \%   &  \protect\cite{RahbordedaneshPolls0319}{\notee}  & R\&B  \\
  $<$  7 June 2009   &  16,500{\notef}    &  61.1\%{\notec}   &  -  &  -  &  27.2\%{\notec}  &  \protect\cite{AlefPolls0318}    & IRIB \\
{$<$  6 June 2009 }   & {16,945 }   & {22.5 \% }  & {~4 \% } & {~7.5 \% }   & {64 \% }  & {\protect\cite{BaznevisPolls0316}}{\noteg}    &  - \\
  $<$  3 June 2009    &  -    &  29.5 \%   &  25.2 \%  &  ~7.5 \%    &  37.5 \%   &  \protect\cite{RahbordedaneshPolls0319}{\notee}    &  R\&B \\
  $<$  1 June 2009  &  -   &  53 \%   &  -  &  -  &  36 \%   &  \protect\cite{presstv2pollsJune1}    &  Raja \\
  26 May--5 June 2009{\noteh}    &  300,000    &  24.61 \%   &  10.14 \%  &  10.72 \%    &  54.53 \%   &  \protect\cite{Ilna}    &  - \\
 31 May 2009    &  -  &  32 \%   &  27 \%  &  ~6 \%    &  36 \%   &  \protect\cite{tabnak13880310}{\notee}    &  R\&B  \\
  $<$  31 May 2009  &   77,058   &  33 \%   &  27 \%  &  ~3 \%    &  36 \%   &  \protect\cite{tabnak13880310}{\notee}    & Baz. \\
 {$<$  27 May 2009 }   &{1650 }  & {35 \% }  &  -  &  -  &{54 \% }  & {\protect\cite{GhalamnewsPolls0306}}{\noteg}    &  - \\
  $<$  26 May 2009  &  -  &  34 \%   &  -  &  -  &  38 \%   &  \protect\cite{presstv_ayandeh_pre26may}    &  Ayan. \\
 5 May 2009    &  -   &  38.1 \%   &  14.7 \%  &  11.1 \%   &  30.6 \%   &  \protect\cite{RahbordedaneshPolls0319}{\notee}    &  R\&B \\
 3--4 May 2009  & -   &  59 \%   &  -  &  -  &  22 \%   &  \protect\cite{presstv2pollsMay12}     &  Raja \\
  $<$  3 May 2009  &  -  &  54 \%   &  -  &  -  &  22 \%    &  \protect\cite{presstv2pollsMay12}    &  gov. \\
 4 April 2009    &  -   &  40.4 \%   &  ~0.6 \%  &  ~7.9 \%   &  23.5 \%   &  \protect\cite{RahbordedaneshPolls0319}{\notee}    &  R\&B \\
 5 March 2009    &  -   &  43.7 \%   &  ~0 \%  &  ~7.2 \%   &  12.9 \%   &  \protect\cite{RahbordedaneshPolls0319}{\notee}    &  R\&B \\
 \hline
    \end{tabular} %% $$ 
    \\ 
      \tabnote{{\notea}The dates of polls with upper
      limits to the polling date are interpreted to be one day before
      the upper limit implicit in the source.}
      \tabnote{{\noteb}Polling organisation if known: 
      Ayandeh News (Ayan.),
      Baznevis (Baz.),
      ``government'' (gov.),
      Islamic Republic of Iran Broadcasting (IRIB),
      Rahbord \& Danesh (R\&B), 
      Rajanews (Raja).}
      \tabnote{{\notec}The source values for Tehran and the countryside are
      combined in the proportions of the official results, i.e. 
      4,179,188:35,066,803 \protect\citep{MOI09}.}
      \tabnote{{\noted}The source states 54--57\%; the mean 55.5\% is adopted here.}
      \tabnote{{\notee}Poll includes internet component. The polls of 5 March, 4
      April, and 5 May also estimate 36.2\%, 27.6\%, and 5.3\% support,
      respectively, for candidate Kh, who announced his withdrawal on 16
      March.}
      \tabnote{{\notef}The source states 16,000--17,000.}
      \tabnote{{\noteg}Source no longer online and no 
      archival copy is known.}
      \tabnote{{\noteh}The median date (31 May) of the ILNA
      poll is adopted for that poll.}
\end{table*}
} %\newcommand\tprepolls

\newcommand\tprepollsprob{
\begin{table*}
  \caption{Internal consistency 
    (confidence level $p < p_{\chi^2}$) 
    among subsets of the pre-election polls and external consistency
    (confidence level $p < p_{y_j}$) of official result $y_j/\sum_j y_j$ with linear
    least-squares equal-weighted best fit to poll evolution $w_{ij}(t_i)$.
  \label{t-prepolls-prob}}
    $$
    \begin{array}{c c c c  r  lllll  lllll} \hline
      \multicolumn{4}{c}{\mbox{selection} } & 
      &
      \multicolumn{5}{c}{ p_{\chi^2} \mbox{(internal)}} &
      \multicolumn{5}{c}{ p_{y_j} \mbox{(external)}} \\
      \mbox{i.}{\notea} & \mbox{u.}{\noteb} & 
      \mathrm{A}^+{\notec} & \mathrm{A}^-{\noted} &
      N_{\mathrm{p}}{\notee} &
      \mathrm{A} & \mathrm{R} & \mathrm{K} & \mathrm{M} & \mathrm{RKO}{\notef} &
      \mathrm{A} & \mathrm{R} & \mathrm{K} & \mathrm{M} & \mathrm{RKO}{\notef} \\
      \hline
      \rule{0ex}{2.5ex}   % strut 
      \mathrm{Y} & \mathrm{Y} & \mathrm{Y} & \mathrm{Y} &  17 &        0 & 
      10^{-11} & 
      0.736 &        0 &        - &    
      10^{-16} & 10^{-11} & 10^{-12} & 10^{-5} &       -   \\
      \mathrm{N} & \mathrm{Y} & \mathrm{Y} & \mathrm{Y} &  10 &        0 &        - &        - &        0 
      &   0.064 &    
      10^{-4} &        - &        - &   0.003 & 10^{-4}    \\
      \mathrm{Y} & \mathrm{N} & \mathrm{Y} & \mathrm{Y} &  15 &        0 & 
      10^{-4} &
      0.638 & 
      10^{-15} &
      - &    
      10^{-11} & 0 & 10^{-8} & 10^{-3} & - \\
      \mathrm{N} & \mathrm{N} & \mathrm{Y} & \mathrm{Y} &   8 &        0 &        - &        - &        0 &   0.064 &
      0.004 &        - &        - &   0.043 & 10^{-4}    \\
      \mathrm{N} & \mathrm{N} & \mathrm{Y} & \mathrm{N} &   5 &    0.273 &        - &        - &    0.153 &   0.948 &  
       0.131 &        - &        - &    0.604 & 10^{-4}   \\
      \mathrm{N} & \mathrm{N} & \mathrm{N} & \mathrm{Y} &   3 &    0.447 &        - &        - &   
      0.050 &    
      0.706 &       
      0 &        - &        - &  0.001 & 10^{-3}   \\
      \hline
    \end{array}$$
    \tabnote{{\notea}Are the partially internet-based polls included (Yes/No)?}
    \tabnote{{\noteb}Are the publicly unarchived polls included?}
    \tabnote{{\notec}Are polls where $w_{i{\mathrm A}} \ge 50\%$ included?}
    \tabnote{{\noted}Are polls where $w_{i{\mathrm A}} < 50\%$ included?}
    \tabnote{{\notee}Number of polls in subset.}
    \tabnote{{\notef}The sum of votes for R, K, and other (O) are inferred
    from the votes for A and M in the cases where the partially internet-based
    polls are excluded. The official result in this case is the sum of votes
    for R and K.}
\end{table*}
} %\newcommand\tprepollsprob

\newcommand\tprepollsbest{
\begin{table*}
  \caption{Expected values E$(w_{ij}^*)$ and standard errors SE$(w_{ij}^*)$ of 12 June vote
    proportions $w_{ij}^* := w_{ij}(t=\mbox{12 June})$
    for candidates $j=$A and M 
    implied by linear least-squares equal-weighted
    best fits to $w_{ij}(t_i)$.
    \label{t-prepolls-best}}
  $$
  \begin{array}{c c c c  cc cc cc cc } \hline 
    \rule{0ex}{2.5ex}   % strut     
    \mbox{i.}{\notea} & \mbox{u.}{\noteb} & 
    \mathrm{A}^+{\notec} & \mathrm{A}^-{\noted} &
    \mathrm{E}(w_{i\mathrm{A}}^*) & 
    \mathrm{SE}(w_{i\mathrm{A}}^*) & 
    \mathrm{E}(w_{i\mathrm{K}}^*) & 
    \mathrm{SE}(w_{i\mathrm{K}}^*) & 
    \mathrm{E}(w_{i\mathrm{M}}^*) & 
    \mathrm{SE}(w_{i\mathrm{M}}^*) &
    \mathrm{E}(w_{i\mathrm{RKO}}^*){\notee}{\notef} & 
    \mathrm{SE}(w_{i\mathrm{RKO}}^*){\notef}
 \\ 
\hline
\mathrm{Y} & \mathrm{Y} & \mathrm{Y} & \mathrm{Y} & 36.4 \% &   3.2 \% &   7.1 \% &   0.9 \% &   44.5 \% &   2.5 \% & - & -  \\
\mathrm{N} & \mathrm{Y} & \mathrm{Y} & \mathrm{Y} &  36.3 \% &   6.5 \% &  -  &  - &   50.1 \% &   5.5 \%  &    13.6 \% &   2.4 \%   \\
\mathrm{Y} & \mathrm{N} & \mathrm{Y} & \mathrm{Y} &  38.4 \% &   3.5 \% &   7.0 \% &   1.1 \% &   40.8 \% &   2.1 \%   & - & -  \\
\mathrm{N} & \mathrm{N} & \mathrm{Y} & \mathrm{Y} &  41.1 \% &   7.4 \% &  -  &  - &   44.2 \% &   5.2 \%   &  14.7 \% &   2.8 \%   \\
\mathrm{N} & \mathrm{N} & \mathrm{Y} & \mathrm{N} & 58.9 \% &   2.3 \% &   -  &  - &   32.3 \% &   2.7 \%    &  8.7 \% &   1.2 \%   \\
\mathrm{N} & \mathrm{N} & \mathrm{N} & \mathrm{Y} &  18.7 \% &   4.6 \% &  -  &  - &   62.8 \% &   8.9 \%   &  18.5 \% &   4.3 \%   \\
\hline
              \end{array}$$
  \tabnote{{\notea}{\noteb}{\notec}{\noted}{\notef}As per Table~\protect\ref{t-prepolls-prob}.}
  \tabnote{{\notee}Official result \protect\citep{MOI09}: $y_{\mathrm{RK}} = 2.5\%$.}
\end{table*}
} %newcommand\tprepollsbest

\newcommand\tprepollsbestlinear{
\begin{table*}
  \caption{\protect\postrefereechanges{Expected values E$(w_{ij}^*)$
      and standard errors SE$(w_{ij}^*)$ of 12 June vote proportions,
      as for Table~\protect\ref{t-prepolls-best}, but for 10,000
      numerical realisations, using the more accurate method (leading
      to smaller standard errors) detailed in
      Sect.~\ref{s-quadratic}.}
    \label{t-prepolls-best-linear}}
  $$
  \begin{array}{c c c c  cc cc cc cc } \hline 
    \rule{0ex}{2.5ex}   % strut     
    \mbox{i.}{\notea} & \mbox{u.}{\noteb} & 
    \mathrm{A}^+{\notec} & \mathrm{A}^-{\noted} &
    \mathrm{E}(w_{i\mathrm{A}}^*) & 
    \mathrm{SE}(w_{i\mathrm{A}}^*) & 
    \mathrm{E}(w_{i\mathrm{K}}^*) & 
    \mathrm{SE}(w_{i\mathrm{K}}^*) & 
    \mathrm{E}(w_{i\mathrm{M}}^*) & 
    \mathrm{SE}(w_{i\mathrm{M}}^*) &
    \mathrm{E}(w_{i\mathrm{RKO}}^*){\notee}{\notef} & 
    \mathrm{SE}(w_{i\mathrm{RKO}}^*){\notef}
 \\ 
\hline
\mathrm{Y} & \mathrm{Y} & \mathrm{Y} & \mathrm{Y} &  36.4 \% &   0.6 \% &     7.1 \% &   0.4 \% &   44.5 \% &   0.6 \%    & - & - \\ 
\mathrm{N} & \mathrm{Y} & \mathrm{Y} & \mathrm{Y} &  36.3 \% &   1.0 \% &    -      &    -      &   50.1 \% &   1.0 \%  &   13.6 \% &   0.9 \%   \\ 
\mathrm{Y} & \mathrm{N} & \mathrm{Y} & \mathrm{Y} &  38.4 \% &   0.7 \% &     7.0 \% &   0.4 \% &   40.8 \% &   0.7 \%    & - & - \\ 
\mathrm{N} & \mathrm{N} & \mathrm{Y} & \mathrm{Y} &  41.1 \% &   1.2 \% &    -      &    -      &   44.2 \% &   1.1 \%   &  14.7 \% &   1.0 \%   \\ 
\mathrm{N} & \mathrm{N} & \mathrm{Y} & \mathrm{N} &  58.9 \% &   1.7 \% &    -      &    -      &   32.3 \% &   1.2 \%   &   8.7 \% &   1.1 \%   \\ 
\mathrm{N} & \mathrm{N} & \mathrm{N} & \mathrm{Y} &  18.7 \% &   2.0 \% &    -      &    -      &   62.8 \% &   2.9 \%   &  18.5 \% &   2.7 \%   \\ 
\hline
              \end{array}$$
  \tabnote{{\notea}{\noteb}{\notec}{\noted}{\notee}{\notef}As per Table~\protect\ref{t-prepolls-best}.}
\end{table*}
} %newcommand\tprepollsbestlinear

\newcommand\tprepollsbestquadratic{
\begin{table*}
  \caption{\protect\postrefereechanges{Expected values E$(w_{ij}^*)$ and standard errors SE$(w_{ij}^*)$ of 12 June vote
    proportions, as for Table~\protect\ref{t-prepolls-best}, but for 10,000 
    numerical realisations and a quadratic least-squares best fit, as detailed in Sect.~\ref{s-quadratic}.}
    \label{t-prepolls-best-quadratic}}
  $$
  \begin{array}{c c c c  cc cc cc cc } \hline 
    \rule{0ex}{2.5ex}   % strut     
    \mbox{i.}{\notea} & \mbox{u.}{\noteb} & 
    \mathrm{A}^+{\notec} & \mathrm{A}^-{\noted} &
    \mathrm{E}(w_{i\mathrm{A}}^*) & 
    \mathrm{SE}(w_{i\mathrm{A}}^*) & 
    \mathrm{E}(w_{i\mathrm{K}}^*) & 
    \mathrm{SE}(w_{i\mathrm{K}}^*) & 
    \mathrm{E}(w_{i\mathrm{M}}^*) & 
    \mathrm{SE}(w_{i\mathrm{M}}^*) &
    \mathrm{E}(w_{i\mathrm{RKO}}^*){\notee}{\notef} & 
    \mathrm{SE}(w_{i\mathrm{RKO}}^*){\notef}
 \\ 
\hline
\mathrm{Y} & \mathrm{Y} & \mathrm{Y} & \mathrm{Y} &  33.3 \% &   0.9 \% &     5.7 \% &   0.6 \% &   46.1 \% &   0.9 \%    & - & - \\ 
\mathrm{N} & \mathrm{Y} & \mathrm{Y} & \mathrm{Y} &  49.4 \% &   1.8 \% &    -      &    -      &   38.6 \% &   1.8 \%    & 12.0 \% &   1.6 \%   \\ 
\mathrm{Y} & \mathrm{N} & \mathrm{Y} & \mathrm{Y} &  35.7 \% &   1.0 \% &     5.4 \% &   0.6 \% &   42.0 \% &   0.9 \%    & - & - \\ 
\mathrm{N} & \mathrm{N} & \mathrm{Y} & \mathrm{Y} &  54.0 \% &   2.1 \% &    -      &    -      &   36.0 \% &   1.9 \%    & 10.0 \% &   1.8 \%   \\ 
\mathrm{N} & \mathrm{N} & \mathrm{Y} & \mathrm{N} &  63.7 \% &   3.3 \% &    -      &    -      &   25.0 \% &   2.4 \%    & 11.3 \% &   2.0 \%   \\ 
\mathrm{N} & \mathrm{N} & \mathrm{N} & \mathrm{Y} &  \mathrm{N/A}^{\noteg} \\
\hline
              \end{array}$$
  \tabnote{{\notea}{\noteb}{\notec}{\noted}{\notee}{\notef}As per Table~\protect\ref{t-prepolls-best}.}
  \tabnote{{\noteg}Too few polls in this case.}
\end{table*}
} %newcommand\tprepollsbestquadratic

\newcommand\tprepollsbestcubic{
\begin{table*}
  \caption{\protect\postrefereechanges{Expected values E$(w_{ij}^*)$ and standard errors SE$(w_{ij}^*)$ of 12 June vote
    proportions, as for Table~\protect\ref{t-prepolls-best}, but for 10,000 
    numerical realisations and a cubic least-squares best fit, as detailed in Sect.~\ref{s-quadratic}.}
    \label{t-prepolls-best-cubic}}
  $$
  \begin{array}{c c c c  cc cc cc cc } \hline 
    \rule{0ex}{2.5ex}   % strut     
    \mbox{i.}{\notea} & \mbox{u.}{\noteb} & 
    \mathrm{A}^+{\notec} & \mathrm{A}^-{\noted} &
    \mathrm{E}(w_{i\mathrm{A}}^*) & 
    \mathrm{SE}(w_{i\mathrm{A}}^*) & 
    \mathrm{E}(w_{i\mathrm{K}}^*) & 
    \mathrm{SE}(w_{i\mathrm{K}}^*) & 
    \mathrm{E}(w_{i\mathrm{M}}^*) & 
    \mathrm{SE}(w_{i\mathrm{M}}^*) &
    \mathrm{E}(w_{i\mathrm{RKO}}^*){\notee}{\notef} & 
    \mathrm{SE}(w_{i\mathrm{RKO}}^*){\notef}
 \\ 
\hline
\mathrm{Y} & \mathrm{Y} & \mathrm{Y} & \mathrm{Y} &  35.5 \% &   1.4 \% &     4.8 \% &   1.0 \% &   43.8 \% &   1.3 \%    & - & - \\ 
\mathrm{N} & \mathrm{Y} & \mathrm{Y} & \mathrm{Y} &  57.8 \% &   3.7 \% &    -      &    -      &   28.5 \% &   3.2 \%    & 13.7 \% &   2.9 \%   \\ 
\mathrm{Y} & \mathrm{N} & \mathrm{Y} & \mathrm{Y} &  37.7 \% &   1.5 \% &     4.2 \% &   1.1 \% &   41.2 \% &   1.3 \%    & - & - \\ 
\mathrm{N} & \mathrm{N} & \mathrm{Y} & \mathrm{Y} &  54.5 \% &   3.7 \% &    -      &    -      &   31.5 \% &   3.1 \%   &  14.0 \% &   2.9 \%   \\ 
\mathrm{N} & \mathrm{N} & \mathrm{Y} & \mathrm{N} &  68.2 \% &   4.8 \% &    -      &    -      &   24.5 \% &   3.3 \%    &  7.3 \% &   3.2 \%   \\ 
\mathrm{N} & \mathrm{N} & \mathrm{N} & \mathrm{Y} &  \mathrm{N/A}^{\noteg} \\
\hline
              \end{array}$$
  \tabnote{{\notea}{\noteb}{\notec}{\noted}{\notee}{\notef}As per Table~\protect\ref{t-prepolls-best}.}
  \tabnote{{\noteg}Too few polls in this case.}
\end{table*}
} %newcommand\tprepollsbestcubic

\newcommand\tKseventies{
\begin{table}
\caption{Frequencies $N$ of the two-digit votes for K starting with
  the digit 7.
\label{t-Kseventies}}
$$\begin{array}{c  r r r r r   r r r r r} \hline 
\rule{0ex}{2.5ex}
{v_{iK}} & 70 & 71 & 72 & 73 & 74 & 75 & 76 & 77 & 78 & 79 \\
      N & 1  & 2  &  1 &  4 &  1 &  4 &  1 &  2 &  1 &  3 \\
\hline 
\end{array}$$
\end{table}
} %\newcommand\tKseventies

\newcommand\tKsevenall{
\begin{table}
\caption{Parity frequencies $N$ 
of the K7-selected vote counts 
and the cumulative binomial tail probabilities $p$.
\label{t-K7-all}}
$$\begin{array}{c  r r r r } \hline 
\rule{0ex}{2.5ex}
& \multicolumn{4}{c}{\mbox{candidate}\; j}  \\
\mbox{selection} & \mathrm{A} & \mathrm{R} & \mathrm{K} & \mathrm{M} \\
\hline
N(v_{ij} \; \mbox{even} \;\vert\; d_1(v_{iK}) =7 ) & 18 & 15 & 18 & 16 \\
N(v_{ij} \; \mbox{odd} \;\vert\; d_1(v_{iK}) =7 ) & 23 & 26 & 23 & 25 \\
p & 0.266 & 0.059 & 0.266 & 0.106 \\
\hline 
\end{array}$$
\end{table}
} %\newcommand\tKsevenall

\newcommand\tpollobservers{
\begin{table}
\caption{Numbers of polling stations $N^{\mathrm poll}$ and polling
  station observers per candidate $N^{\mathrm poll}_{j}$
  \protect\citep{Brill10}$^{\mathrm a}$ and an advisor for M
  \protect\citep{Lucas09}$^{\mathrm b}$; ``-'' indicates no value claimed.
  \label{t-pollobservers}}
$$\begin{array}{c  r r r r r } \hline 
  \rule{0ex}{2.5ex}
  N^{\mathrm poll}
  & \multicolumn{4}{c}{N^{\mathrm poll}_{j} } & {\mathrm{ref}} \\
  \mbox{selection} & \mathrm{A} & \mathrm{R} & \mathrm{K} & \mathrm{M} \\
  \hline
  45692 &
  33058 & 5421 & 13506 & 40676 & {}^{\mathrm a} \\
  - & - & - & - & 25000 & {}^{\mathrm b} \\
  \hline 
\end{array}$$
\end{table}
} %\newcommand\tpollobservers

\newcommand\tKseven{
\begin{table}
\caption{Votes for K and proportion of votes for A for the 
six voting areas with the greatest numbers of total votes.
\label{t-Kseven}}
$$\begin{array}{c r r r } \hline 
\mathrm{voting~area} 
& x_i
& v_{i\mathrm K}
& v_{i\mathrm A}/x_i
\rule{0ex}{2.5ex}  \\  % strut 
\hline
\mathrm{ Tabriz} &   876919 & 3513 &0.497  \\
\mathrm{ Shiraz} &   947168 & 7078 &0.600  \\  
\mathrm{ Karaj} &   950243 & 8057 &0.537  \\  
\mathrm{Isfahan} &   1095399 & 7002 &0.609 \\ 
\mathrm{Mashhad} &   1536106 & 7098 &0.669 \\  
\mathrm{Tehran} &   4179188 & 43073 &0.433 \\ 
\hline 
\end{array}$$
\end{table}
} % \newcommand\tKseven
\newcommand\mytitle{A first-digit anomaly in the 2009 Iranian presidential election}

\begin{document}

%\begin{frontmatter}
\doi{10.1080/02664763.YYYY.XXXXXX}
\issn{1360-0532}
\issnp{0266-4763}
%\jvol{00} \jnum{00} \jyear{2012} \jmonth{February}

% "Title of the paper"
\title{\protect\mytitle}
%\runtitle{Iran 2009 election anomaly}
\markboth{B.F. Roukema}{Journal of Applied Statistics}

% indicate corresponding author with \corref{}
% \author{\fnms{John} \snm{Smith}\corref{}\ead[label=e1]{smith@foo.com}\thanksref{t1}}
% \thankstext{t1}{Thanks to somebody} 
% \address{line 1\\ line 2\\ printead{e1}}
% \affiliation{Some University}

\author{Boudewijn F. Roukema\thanks{{\em{Email: boud~astro.umk.pl}}}\\
  \vspace{6pt} {\em Toru\'n Centre for Astronomy, 
    Faculty of Physics, Astronomy and Informatics,
    Nicolaus Copernicus University, ul. Gagarina
    11, 87-100 Toru\'n, Poland}\\
    \vspace{6pt}\received{[received date: \ldots]}}

%\author{\fnms{Boudewijn F.} \snm{Roukema}\ead[label=e1]{boud@astro.umk.pl}}
%\author{Boudewijn F. Roukema}
%\ead{boud@astro.umk.pl}
%\affiliation{Toru\'n Centre for Astronomy, Nicolaus Copernicus University}
%\address{
%Toru\'n Centre for Astronomy, UMK \\
%ul. Gagarina 11, 87-100 Toru\'n, Poland\\
%\printead{e1}
%}
%\and
%\author{\fnms{???} \snm{???}\ead[label=e2]{???}}
%\address{\printead{e2}}
%\affiliation{???}

%\runauthor{B.F. Roukema}

\maketitle

\begin{abstract} %% max 150 words
%WCWC word count
A local bootstrap method is proposed for the analysis of electoral vote-count
first-digit frequencies, complementing the Benford's Law
limit.  The method is calibrated on five presidential-election first rounds
(2002--2006) and applied to the 2009 Iranian presidential-election 
first round.
Candidate K has a highly significant ($p< 0.15\%$) excess of vote counts starting with the digit~7. This 
leads to other anomalies, two of which are individually significant at 
$p\sim 0.1\%$, and one at $p\sim 1\%$.
%Using this as a selection criterion leads to several coincidences, one of
%which selects those three of the six most populous voting areas that
%have the greatest proportions of votes for candidate A.
Independently, Iranian pre-election opinion polls significantly reject
the official results unless the
five polls favouring candidate A are considered alone. If the latter
represent normalised data
\protect\postrefereechanges{and a linear, least-squares, equal-weighted
fit is used,}
%(undecided/other responses excluded), 
then 
either candidates R and K suffered a sudden, dramatic 
($70\%\pm 15\%$) loss of electoral support just prior to the election, 
% during the last several days preceding the election, 
%or the official vote results for R and/or K are 
or the official results 
are rejected ($p\sim 0.01\%$). 
\end{abstract}

%% http://www.ams.org/mathscinet/msc/msc.html?t=62Pxx&btn=Current
%\begin{keyword} 
%[class=AMS]
%\kwd[Primary]{62P25}  % 62P25    	Applications to social sciences
%\kwd{}
%\kwd[; secondary ]{}
%\end{keyword}

\begin{keywords}
  {presidential election} ---
  {Benford's Law} ---
  {bootstrap} ---
  {Iran}
\end{keywords}

\begin{classcode}
  62P25
% MSC2010:	62P25    	Applications to social sciences
% http://www.ams.org/mathscinet/msc/msc2010.html?t=62Pxx&btn=Current
\end{classcode}

%\end{frontmatter}

%MAINTEXT word count

\section{Introduction} \label{s-intro}

The results of the 12 June 2009 presidential-election first round held in the
Islamic Republic of Iran are of high political importance in
Iran. International interest in these results is also considerable.
On 14 June 2009, the Ministry of the Interior (MOI) published a table
of the numbers of votes $v_{ij}$ received by the $j$-th of the $n=4$
candidates for the $i$-th of $m=366$ voting areas \citep{MOI09}. In
order to avoid focussing on personalities, the four candidates will be
referred to here as A, R, K, and M, following the order given in the
table. These letters correspond to the conventional Roman alphabet
transliteration of the four candidates' names by which they are
frequently referred to. The total votes $y_j := \sum_i v_{ij}$ for
these four candidates from the MOI table give A as the winner with
24,515,209 votes, against R with 659,281 votes, K with 328,979 votes,
and M with 13,225,330 votes.
%with 24515209 votes, 
%against R with 659281 votes, 
%K with  328979 votes, and 
%M with 13225330 votes.
A second voting round was not held.

The total numbers of votes per voting area $x_i \ge \sum_j v_{ij}$
(invalid votes are included in $x_i$) in the MOI's data vary from
about $10^4$ to $10^6$, i.e. by two orders of magnitude (powers of
10).  This suggests that Benford's Law (\citealt{Newc1881}; \citealt{Benf38}) may be
applicable, i.e.  it may be useful to test the null hypothesis that
the first digit in the candidates' absolute numbers of votes,
represented in the decimal system, are consistent with random
selection from a uniform, base 10 logarithmic distribution modulo
1. This test has been historically proposed for finding hints of
artificial interference in statistical data sets.  The reason is that
common intuition suggests that the frequencies of occurrence of the
first digit (using the decimal system) in a large data set of
empirical data that cover at least an order of magnitude should be
approximately equally distributed between the nine non-zero digits, i.e. 
about $1/9 \approx 11\%$ of the first digits should be 1,
about $11\%$ should be 2, \ldots, 
and about $11\%$ should be 9.
Benford's Law contradicts this. For example, the most striking
characteristic of Benford's Law follows from
Eq.~(\ref{e-benfordbasic}) below: the first digit is 1 with a
frequency of $\log_{10}2 \approx 30\%$, i.e. it occurs much more
frequently than any other digit, with a much higher frequency than
$11\%$.  The frequencies of other digits than the first digit can also
be analysed with Benford's Law.  Independently of the present
analysis, the distributions of the {\em second} digit were analysed
for this same data set \citep{Meb09a} and the {\em last} digits of a
related data set also released by the MOI were analysed
\citep{BebScacco} using a method developed earlier using Nigerian electoral data
\citep{BebScacco08}. 

The decimal first digit version of Benford's Law
\citep{Newc1881,Benf38} can be given informally by stating that for
many real world samples of values that span several orders of
magnitude, the relative frequency of the occurrence of digit $d \in
\{1, \ldots, 9 \}$ as the first digit in decimal representations of
real numbers tends towards
\begin{equation}
f(d) = \log_{10}\left(1+\frac{1}{d}\right)
\label{e-benfordbasic}
\end{equation}
as the sample size and logarithmic distribution width increase.
Equation~(\ref{e-benfordbasic}) should be a good approximation if
a sample can be expected to 
be drawn from a probability distribution of a random variable $X$ 
that varies slowly enough over
several orders of magnitude in such a way that 
\begin{equation}
Y := \log_{10} X - \lfloor \log_{10} X \rfloor , 
\label{e-benfordfolded}
\end{equation}
i.e. the {\em folding} of $X$ to a single decade, is approximately uniform,
where $\lfloor x \rfloor$ is the greatest integer $ \le x$. Because
the logarithmic scale is folded into a single decade, statistical
populations do not necessarily need to span many orders of magnitude
in order to approximately satisfy Benford's Law.  
A folded logarithmic distribution 
(for a fixed base) 
will tend towards a uniform distribution as the logarithmic variance 
of the unfolded distribution increases.
However, it is not clear how fast and smooth this convergence occurs.
For example, a logarithmically uniform distribution on $z_1 \le \log_{10}
X < z_2 $ will not fold to a uniform distribution unless $z_2 - z_1$
is an integer.  Nevertheless, historically, Benford's Law has been
found to apply well in practice in many cases, especially when several
different distributions are combined to 
constitute a single distribution.

A detailed justification and generalisation of several equivalent
formulations of Benford's Law are given by Hill, who refers to
Benford's Law as the ``central-limit-like theorem for significant digits'' 
\citep{Hill95}.  To what degree should Benford's Law apply to
the Iranian MOI data set for the 2009 presidential election first
round of voting? To the extent that the total voting populations $x_i$
of the voting areas used for the MOI data set constitute a mix of many
processes---the growth of towns and cities over thousands of
years---that can be modelled as if they were a mix of several
different (mostly non-uniform)
random processes corresponding to a mix of probability
distributions, including the still rapid (over 1\%) population growth
and the freedom to vote anywhere in Iran independently of one's place
of residence, these total votes $x_i$ may to some degree satisfy
base-neutrality and/or scale-neutrality. [See Definition 5 of
  \protect\cite{Hill95} for definitions of base and scale bias.] On
the other hand, political and administrative effects might introduce
strong base or scale biases.

However, rather than $x_i$, what are even more likely to constitute a
mix of (mostly non-uniform) random processes are the processes leading
to the fractional voting rates $\vijhat := v_{ij} / x_i \le 1$ for
individual candidates for voting areas of a given 
population size $x_i$, where approximately equal $x_i$'s are grouped 
together at $x$. If (i) {\em the voting rates $\vijhat$ vary by a
  large fraction of an order of magnitude near $x_i \approx x$,} then
the votes $v_{ij}$ in the voting areas of size $x_i \approx x$ should
also scatter by a large fraction of an order of magnitude.  If, in
addition, (ii) {\em the distribution of $\log_{10} \vijhat(x_i \approx x)$ varies
  slowly enough with $\log_{10} x$,} then it should be possible that
scale-neutrality is satisfied well enough for Benford's Law to be
applicable.

Iranian society has a continuous and complex urban history dating back
about 6000 years.
%While the beliefs in social homogeneity that exist in some parts of
%the world may be culturally constructed myths of national identity
%rather than sociological reality, there are few claims of
%sociological homogeneity in Iran.
Hence, at least as much as in any other complex society, Iranian
voting patterns can be expected to be the result of a highly rich mix
of numerous social, political, economic, and historical factors that can
influence voting decisions, as well as 
\postrefereechanges{natural environmental factors (e.g. the
weather \citep{Gomez07weather,AchenBartels04})} and the cognitive margin of freedom that individuals retain with
respect to these broader factors.  Are all these factors enough to satisfy
the conditions (i) and (ii) sufficiently well 
%that the voting rates $\vijhat$ vary by a large
%fraction of an order of magnitude at a given $x$, and (ii) that the
%dependence of the distribution of $\log_{10} \vijhat$ at a given
%approximate $x$ on $\log_{10} x$ is weak, 
such that the MOI data set can
be statistically tested using Benford's Law for the vote counts
$v_{ij}$?

These qualitative descriptions of the two conditions hide a lot
of quantitative detail, which could require as many as six parameters 
to quantify. 
% For example, these could include (i) how many multiples
% of the standard deviation correspond to ``vary'', how much is ``large'',
% how close to $x$ is considered to satisfy ``approximately $x$'', (ii) 
% how much is ``approximately $x$'' (again), at least two
% parameters should characterise the distribution of $\log_{10} \vijhat$,
% and how much is ``weak''. 
For example, these could include, respectively, quantitative definitions of (i)
``vary'' in terms of standard deviations, ``large'', and ``$\approx$'', and of
(ii) at least two parameters to characterise the
distribution of $\log_{10} \vijhat$, and ``slowly''.  Clearly, it
would be preferable to use a non-parametric alternative. Moreover,
rather than attempt a theoretical proof of whether or not a
formalisation of these conditions would imply Benford's Law, it would
be preferable to formalise the conditions in such a way that they
define a statistical model of the data that can be used directly, independently
of how closely the data can be expected to approach the limiting form
of Benford's Law. 

Hence, here, a statistical model is defined using a local bootstrap
method, motivated by conditions (i) and (ii).  This is presented in
Section~\ref{s-method-boot}. This method is non-parametric in terms of
the data set being tested, but uses one fixed, base-related
parameter. The aim of the ``local'' property of the method is to
increase the method's specificity (reduce the chance that it falsely
rejects the null hypothesis that no artificial interference in the
data has taken place), at the cost of decreasing the method's power
(the chance that the method misses genuine anomalies is
increased). Bootstrap methods can give biased and/or skewed estimates of confidence
intervals, so the probabilities inferred using the method are
calibrated against several recent presidential-election first rounds prior to
the 2009 Iranian presidential-election first round.  These data sets and the
Iranian 2009 MOI vote counts are described in
Section~\ref{s-datasets}.  The first digit frequencies in the Iranian
2009 MOI data are examined using this calibrated probability distribution.

In addition to the internal statistical test provided by first-digit
frequencies, external tests of the MOI data can be made by
comparing these data to a compilation of pre-election opinion polls
that were openly and transparently documented by the English-language community 
prior to the election day. This compilation of polls
is described in Section~\ref{s-prepolls-data}.

The calibration of the local bootstrap simulations 
using the earlier presidential elections is presented in 
Section~\ref{s-calib-res}.
The first-digit frequencies of the four candidates' official vote counts
and their raw and calibrated confidence levels are presented in 
Section~\ref{s-BL-res} and 
Figs~\ref{f-Aconflev}--\ref{f-Mconflev}.
%Fig.~\ref{f-Aconflev}.
The internal consistency among the pre-election opinion polls and their
consistency with the MOI data set are presented in Section~\ref{s-prepolls-res}.
Discussion, including
several anomalies that follow from the basic result, is given in
Section~\ref{s-disc}.
\postrefereechanges{A summary and conclusion} 
are presented in Section~\ref{s-conc}.

\section{Method} \label{s-method}

\subsection{Data and model files} \label{s-files}
%A plain text {\sc octave} script \citep{Rouk09suppB} is available 
%for making these calculations.
The data from \cite{MOI09} used in this analysis and the data from the
French (2002, 2007), Iranian (2005), Polish (2005) and Brazilian
(2006) presidential-election first rounds are listed in the text file {\sc
  pres\_elections}. An {\sc octave} script\footnote{\url{http://www.gnu.org/software/octave}}
 is provided as a plain text
file {\sc benford\_elections.m} for carrying out the analysis in this
paper, using the input file {\sc pres\_elections}. Both files are part
of the preprint version of this article at ArXiv.org.\footnote{\protect\url{http://arXiv.org/e-print/0906.2789}}

\subsection{Empirical, 
  local bootstrap model of presidential election vote counts}
\label{s-method-boot}

An empirical, local bootstrap model of presidential election
vote counts is defined as follows.

\subsubsection{Local bootstrap}

For the 2009 Iranian presidential election or another similar
presidential election, let the numbers of votes received by the $j$-th
of the $n$ candidates for the $i$-th of $m$ voting areas be $v_{ij}$.
Let the total numbers of votes per voting area be $x_i :=
\sum_{j=1}^{n+1} v_{ij} $ (where $v_{i(n+1)}$ represents invalid
votes if known, and otherwise is set to zero), 
the total votes per candidate be $y_j := \sum_i v_{ij}$, and
the normalised voting rates (estimates of a candidate's electoral
popularity in a given area) be $\vijhat := v_{ij} / x_i \le 1$. 

Sort the $x_i$ such that the sequence $x_{i_1}, x_{i_2}, \ldots,
x_{i_m}$ is (w.l.o.g.)  in ascending order.
\begin{definition}
  A local bootstrap realisation for candidate $j$ is a
  set of simulated votes 
  \begin{equation}
    \{ \viprimejhat \, x_{i_k} \}_{k=1,\ldots,m} 
    \label{e-viprimejhat}
  \end{equation}
  where the values $k'(k)$
  are drawn randomly from 
  \postrefereechanges{$G$, a realisation of 
    a Gaussian probability density function of
    width $m\Delta$, where}
  $\Delta := \log_{10}(10/9) \approx 0.0458$, and centred at $k$,
  truncated at the limits of $\{k\}$,
  i.e.
  \begin{equation}
    k'(k) := \max(1,\, \min(m,\, \lfloor G(k, m \Delta) + 0.5 \rfloor )).
  \end{equation}
  \label{d-local-boot}
\end{definition}
\postrefereechanges{The meaning of $\viprimejhat$ can be thought of by
  starting with an extremely small value of $\Delta$, rather than the
  defined value.  For example, $\Delta = 10^{-100}$ would in practice
  make all realisations of $G(k,m\Delta)$ on a 2012 computer yield
  $G = k$, so that, effectively, $k'(k) \equiv k$ and $\viprimejhat \equiv
  \vikjhat := v_{{i_k}j} / x_{i_k}$, giving a realisation identical to
  the empirical data set. Increasing $\Delta$ to 0.0458
  allows a bootstrap from voting areas with the index $k'$ slightly
  lower or higher than $k$, i.e. a ``real'' voting rate for 
  candidate $j$ is selected from a voting area with a slightly 
  lower or higher (or possibly the same) total voting population. 
  Thus, this is a ``local'' bootstrap---the simulated data are
  generated using the empirical data directly, but instead of
  the standard bootstrap method, the random selection process
  consists of just a slight ``nudge'' to the original data among
  voting areas of about the same voting population size.}

If the $x_{i_k}$ are distributed log-uniformly over one decade,
i.e. if $\log_{10} x_{i_k}$ is distributed log-uniformly within the
decade from $z_1$ to $10z_1$ for some $z_1$, then the fraction of $x_{i_k}$ 
values contained within an interval in $k$ of width $m\Delta$ 
is $(m\Delta)/m = \Delta$. That is, in this case, the smoothing scale
in $x_{i_k}$ is that of the narrowest first-digit interval, 
$\Delta := \log_{10} (10/9)$. Narrower and wider $x_{i_k}$ distributions
will be less and more smoothed, respectively.

A full bootstrap realisation over the set of voting rates
$\vijhat$ for a given candidate $j$ would be equivalent to
assuming that the probability distributions for voting rates in low
and high population voting areas are independent and identically
distributed, which is unlikely to be realistic. Apart from some
correlation between voting area size $x_{i_k}$ and various
sociological characteristics, the Poisson nature of many processes
should modify the variance of the voting rate as a function of
$x_{i_k}$.

If Definition~\ref{d-local-boot} is extended to 
the limit of $\Delta \rightarrow 0$, i.e. if the ``local'' aspect
of the definition is taken to its limit, 
then the realisations 
approach an exact reproduction of the empirical data set itself.
For a value of $\Delta$ that is small, but not so small as to mostly
generate exact reproductions of the data,
the realisations should
sample the distribution of voting rates
$\vikjhat$ near $x_{i_k}$.
That is, any vote realisation $\{ \viprimejhat x_{i_k} \}_{|k-k_0| < m\Delta}$ 
should approximate the actual distribution $\{ v_{{i_k}j} \}_{|k-k_0| < m\Delta}$ 
near any $x_{i_{k_0}}$. 
Hence, %% for $\Delta := \log_{10} (10/9)$,  %% redundant?
within the smallest of the nine logarithmic intervals relevant for 
decimal first digit counts, the realisations should (a)
(re-)sample the empirical distribution of voting rates near $x_{i_k}$, 
whether or not the variance is large or small, and (b) should
approximately mimic the changing properties of this distribution
as $x_{i_k}$ decreases or increases. That is, the realisations should
formalise the conditions (i) and (ii) presented in Section~\ref{s-intro}.
To the degree that these two conditions are satisfied, Benford's Law
is likely to be a good approximation. However, the 
first-digit frequency distributions inferred from the 
local bootstrap realisations
can be used 
independently of the degree to which the two conditions are satisfied,
i.e., without assuming the Benford's Law limiting case. 

Thus, the voting data $x_i$ and $v_{ij}$ themselves are used to construct
discrete probability distributions from which simulated samples of the
set of votes are generated using random resampling allowing repeats. The
simulated samples are used to estimate the confidence intervals for
first-digit counts.  The method is ``local'' in the sense that each
bootstrap is performed within a small subset of the data near a given
total vote $x_{i_k}$, so that the simulated data set should very closely
match the empirical data set. The method is non-parametric in terms of
the data set, and has a fixed, base-related dependence through
$\Delta$.

For these reasons, the method is conservative. By using the data to
model itself, some types of artificial interference in the data may be
mimicked in the simulations, so that the interference
would not be detected. In contrast, any statistically significant
difference between the simulations and the data would imply that the
data are unusually sensitive to resampling the voting rates 
$\vijhat$ for voting areas of approximately the same voting
population $x$. 

For example, suppose that about 20\% of voters in voting areas of
voting population $x \approx 80,000$ typically voted for candidate
$j$, but this varied down to about 10\% and up to about 40\%,
i.e. from 8,000 to 32,000 people voted for candidate $j$ in 
voting areas that have about 80,000 voters, with 16,000 being a typical
vote for candidate $j$.  The local bootstrap model proposed here assumes that
reselecting these percentages randomly from within the same range, 8,000 to
32,000, for this same set of voting areas, 
and applying the same process for each $x$, 
should not affect the overall statistical results. 
The model does not assume any particular shape of
the distribution of $v_{ij}$ in the range from 8,000 to 32,000,
e.g. normal or log-normal, unimodal or multimodal, nor does it assume any
particular skewness or kurtosis. It only assumes that the distribution should
statistically match the observed distribution of the voting 
rates $\vikjhat$ within this limited subset of voting areas.

If outliers (rare extreme values, e.g.  800 votes for candidate $j$ in
one of the voting areas of 80,000 voters in the example just given)
exist among the $\vijhat$ of the empirical data set, then they
will also be included in the simulated data sets with the same
frequencies (in the limit of many realisations). This is an
example of a potentially suspect feature of the empirical data
set---outliers---that would not be detected 
to be significantly unusual by this method.
Hence, it is difficult to see how randomly reassigning votes in a statistically
identical way among these voting areas should destroy any statistical
properties of the votes, such as their decimal first-digit frequencies.

\subsubsection{Empirical calibration} \label{s-meth-calibration}

Using candidates in presidential elections other than the 2009 Iranian
election first round, as detailed in Section~\ref{s-datasets-other}, 
local bootstrap realisations using Definition~\ref{d-local-boot}
are realised for each candidate $j$, except for candidates whose
votes are too low. Candidates are considered to have vote counts
that are too low if
%%% if 1\% or more of the lowest vote counts are \le 1, then ignore
1\% or more of their vote counts satisfy $v_{ij} \le 1$. 
A vote count of zero has an undefined logarithm, and vote counts below 10 
are likely to create strong discreteness effects. 
Hill's formalisation of Benford's Law strictly applies only to random variables
on the set of positive reals, not to a random variable on non-negative
integers \citep{Hill95}. 
Hence, low vote counts are likely to decrease the degree to which 
Benford's-Law-like behaviour occurs.
Candidate 10 %12-2=10
in the Polish 2005 first-round election and candidates 1 and 4 %[3 6] .-2
in the Brazilian 2006 first-round election are excluded from the analysis
using this definition of ``too low''.
This leaves 51 candidates from the five elections.
None of the Iranian 2009 election first-round candidates are excluded
by this criterion.

From an ensemble of local bootstrap realisations for a given candidate
$j$ in a given election, let the 
\postrefereechanges{relative} frequency
with which
the frequency of first digit $d_1$ in a simulated vote count 
$\{ \viprimejhat x_{i_k}\}_{k=1,\ldots,m}$ 
is found to be below the frequency of $d_1$ in
the candidate's empirical vote count 
$\{v_{ij}\}_{i=1,\ldots,m}$  
be ${\cboot}_j(d_1)$.
% \begin{equation}
%   {\cboot}_j(d_1) := 
%   P\left( \fboot\left(D_1(\viprimejhat x_{i_k}) = d_1\right) 
%   < f\left(D_1(v_{ij}) = d_1\right) \right),
% \label{e-defn-cbootjd}
% \end{equation}
% where 
% \begin{equation}
%   D_1(v_{ij}) :=  \lfloor 
%   10^{\log_{10} v_{ij} - \lfloor \log_{10} v_{ij}\rfloor}
%   \rfloor 
% \end{equation}
% is the (decimal) first-digit function.
This corresponds to an upper, one-sided, bootstrap-estimated confidence
level when ${\cboot}_j > 0.5$.
%Where ${\cboot}_j > 0.5$ or $< 0.5$, 
%this will be used as an upper or lower confidence 
%level, written ${\cboot}^+$ or ${\cboot}^-$,
%respectively, 
%i.e. ${\cboot}^- := {\cboot} \vert_{(0,0.5)},$
% ${\cboot}^+ := {\cboot} \vert_{(0.5,1)}.$
%Midpoints of frequency intervals are used. This 
%excludes exact equality in ${\cboot}^{\pm}_j = 0.5$. 

Bootstrap methods in general can require corrections for bias and skewness.
One recommended way to correct for this in the absence of external empirical
data is the bias-corrected, accelerated method. However, for small
samples this is not always reliable in practice
\citep{SchrWill00}. Moreover, in the present case, we have several 
empirical data sets that are publicly available for analysis, are 
independent of the data set of interest, 
and should represent similar sociological processes.
Hence, the approach adopted here is to use
these empirical data sets to calibrate a correction for bias and skewness.
%Literature reports on accelerated bootstraps 
%\citep{SchrWill00} suggest that for the $c=95\%$
%two-sided confidence level, bootstrap estimates of $1-c$ could be 

The bootstrap confidence levels ${\cboot}_j(d)$ obtained for the 9
non-zero digits for the 51 candidates constitute a set of 
$\nboot=459$
values of $\cboot$. 
If the local bootstrap method is accurate in estimating
$\cboot$, then the frequency distribution of $\cboot$ should be
approximated by $P(c < \cboot) = \cboot$. If this is not the case,
then the actual frequency distribution of $\cboot$ over the whole
control data set of 459 values can be used to estimate an {\em
  empirical} set of confidence levels $\cempir$. This implicitly
assumes that the 459 values are independently drawn from a single
probability distribution. This is clearly only an approximation, 
depending on the degree to which the number of digits (9),
the numbers of candidates per election (5-16, where low vote
candidates have been excluded as stated above), and the number
of elections (5) are high enough. None of these three numbers is
individually high, but putting them together, $\nboot = 459$
may be high enough for this to be a sufficient approximation.
Provided that the
function $\cempir(\cboot)$ is smooth enough over the range of
interest, this can be used to convert individual $\cboot$ estimates
in the data set of interest to more accurate confidence levels
$\cempir(\cboot)$. This is the approach adopted here.

The most important regions of the function $\cempir(\cboot)$ are
those near 0 and 1. In particular, any fit $\cempir(\cboot)$
must satisfy $\cempir(0)=0$ and $\cempir(1)=1$.
Let us define 
\begin{eqnarray}
 {\alphaboot}^- := 1 - \cboot,  && \cboot < \cbootstar \nonumber \\
 {\alphaboot}^+ := \cboot,  && \cboot \ge \cbootstar, 
 \label{e-alphaboot}
\end{eqnarray}
where $\cbootstar$ is the median value of $\cboot$. The $\nboot$ 
values of $\cboot$ are sorted in ascending order and each $k$-th
value is paired with 
\begin{equation}
  \cempir(k) := (k -0.5)/\nboot,
  \label{e-cempir}
\end{equation}
so that 
\begin{eqnarray}
 {\alphaempir}^- := 1 - \cempir,  && \cboot < \cbootstar \nonumber \\
 {\alphaempir}^+ := \cempir,  && \cboot \ge \cbootstar
 \label{e-alphaempir}
\end{eqnarray}
can be defined. Hence, an almost-everywhere 
smooth, continuous, piecewise fit $\cempir(\cboot)$ 
allowing one free parameter for each part of $\cempir(\cboot)$ 
is
obtained by (non-linear) least squares fitting 
$2{\alphaempir}^-$ against 
$({\alphaboot}^- / \cbootstar)^{\beta^-}$ and
$2{\alphaempir}^+$ against 
$[{\alphaboot}^+ / (1-\cbootstar)]^{\beta^+}$, 
over the intervals $\cboot < \cbootstar$ and 
 $\cboot \ge \cbootstar$, respectively, where 
$\beta^-$ and $\beta^+$ are free parameters. The fitted solution
is 
\begin{equation}
\cempir(\cboot) = 
\left\{ 
\begin{array}{l c c}
   0.5 \left( \frac{\cboot}{\cbootstar} \right)^{\beta^-} 
  & , & \cboot < \cbootstar \\
  1 - 0.5 \left( \frac{1-\cboot}{1-\cbootstar} \right)^{\beta^+} 
  & , & \cboot \ge \cbootstar.
\end{array}
\right.
\label{e-cempir-cboot}
\end{equation}
This function necessarily satisfies 
$\cempir(0)=0$ and $\cempir(1)=1$ and is continuous through 
$\cboot = \cbootstar$, 
where $\cempir(\cbootstar) = 0.5$.
The first derivative 
$\mathrm{d} \cempir/\mathrm{d} \cboot$
is not necessarily continuous at
$\cboot = \cbootstar$, but if the values of $\cboot$ are 
themselves distributed smoothly, then the derivative should not change
abruptly.
\postrefereechanges{The non-linear
least squares fitting method used in this work to calculate 
$\beta^-$ and $\beta^+$ (Sect.~\ref{s-calib-res}, Fig.~\ref{f-calib})
is iterative.
The power $\beta^-$ (or $\beta^+$)
is initially assumed to lie in the range $(0.2,5.0)$.
Ten uniformly spaced values in this range are tested and
the minimum of $\chi^2$ is used to chose the best estimate for that
iteration. Successive iterations halve the width of the range,
centre it on the previous best fit, and test ten values in the
new, smaller range. Fifty iterations easily converge and 
give the accuracy required.}

%%% for the ir2009a data set: $p=5\%, 1\%, 0.1\%$

%\fhistfolded

%  c = count09(10.^(log10(per_candid*total) .- floor(log10(per_candid*total))));
%  result = c ./sum(c); # normalise

\subsection{Data sets} \label{s-datasets}

\subsubsection{Iranian Ministry of the Interior 14 June 2009 data set}

\sloppy
The primary data of interest here consist of the data set published by
the Iranian Ministry of the Interior (MOI) data on 14 June 2009
\citep{MOI09}. The data set was archived the same day at the URL
\url{http://www.webcitation.org/5hXHfYNbN}.\footnote{Retrieval of the
archival version may require manual removal of html header text and 
tab symbols, e.g. using  
standard GNU/Linux tools,
{\tt 
tail -n 1236 MOI.bin $\vert$ head -n 1235 $\vert$ 
sed -e 's/\^{}{\textbackslash\textbackslash}t{\textbackslash\textbackslash}t//' 
 $>$ MOI.xls}, where the retrieved file is named {\tt MOI.bin} and the 
output file with head and tail removed is to be named {\tt MOI.xls}.} %% needed in Feb 2010
A plain text form of %the Iranian Ministry of the Interior (MOI) data,
this data, 
along with data from the calibration set of
 presidential elections, is provided as the file {\sc pres\_elections}
(see Sect.~\ref{s-files}).
The following day, the MOI published a new version of the data with
minor changes.
The largest obvious modification
was that the total number of votes in Bandar-Abbas in the 14 June file was 390,141,
exactly 100,000 votes greater than the sum of the individual votes and
invalid votes, and greater than the 2005 Bandar-Abbas population of about 350,000,
while in the 15 June file, this was corrected to 290,141. Here, the original 14 June
file is used. 
These two {\sc xls} format files 
use Western Arabic (European) numerals. 
A third file, in {\sc pdf} format and using 
Arabic-Indic (not Persian) numerals \citep[Tables 8-1, 8-2][]{Unicode520},
was also published on 15 June 2009 \citep{MOI09c}. 
The URL that contained links to each of these
three files is archived at \url{http://www.webcitation.org/5hYWAcdhW}.
% In full this is:
% \url{http://www.moi.ir/Portal/Home/ShowPage.aspx?Object=News&ID=e3dffc8f-9d5a-4a54-bbcd-74ce90361c62&LayoutID=b05ef124-0db1-4d33-b0b6-90f50139044b&CategoryID=832a711b-95fe-4505-8aa3-38f5e17309c9}.

\subsubsection{Control data set: other recent, pre-2009 presidential-election first rounds}
\label{s-datasets-other}

For comparison with the Iranian election, data from the first round of
voting in several comparable, recent elections were gathered.  
Some negative and positive criteria for attempting to select statistically and
sociologically comparable data sets are the following.  
%\begin{list}{(\roman{enumi})}{\usecounter{enumi}}
\begin{list}{(\arabic{enumi})}{\usecounter{enumi}}
\item 
  Parliamentary/congressional elections would differ from the 2009
  Iranian presidential election in that in the former case, voters
  often choose political parties rather than focus on candidates, so
  the types of voting behaviour patterns may not sufficiently match
  those in presidential elections.
\item
  Presidential-election second rounds usually only have two
  candidates, so would restrict the parameter freedom of the random
  variables more than in a presidential election with many
  candidates. Single-round, non-preferential elections also have less
  parameter freedom than runoff or preferential elections, since the
  former discourage voting for minor candidates.
\item
  Countries with low populations risk having a narrower logarithmic
  range of voting area sizes (and, hence, a narrower logarithmic range
  in votes for any given candidate). Also, minority candidates in 
  lowly-populated countries are likely
  to have votes in some areas that tend towards the zero-vote limit
  mentioned above.
\item
  Practical (i.e. online) access to the data is preferable so that
  the data can be independently, publicly archived and
  independent researchers can obtain the (possibly updated) data directly from the
  official sources.
\item
  There should be no major controversies regarding suspected
  artificial interference in the election results.
\end{list}
Table~\ref{t-pres-elections} shows the characteristics of online
presidential-election first rounds from relatively highly populated
countries from different geographical areas that were found for this
study, in addition to those of the Iranian 2009 election. These
include a well-established democracy in Western Europe, two countries
that have had regular elections for a few decades since emerging from
left-wing and right-wing authoritarian political systems, and the
Iranian 2005 first-round presidential election.
\postrefereechanges{The sociological diversity among these countries
  and the time span of sustained electoral politics and elections
  among them should imply conservative results, in the sense that an
  election which is statistically unusual in comparison to this
  calibration set of elections is unusual in comparison to a diverse
  set of sociological situations rather than in comparison to a narrow
  set of sociological situations.}
The administrative
division levels 
\postrefereechanges{(called ``voting areas'' in this paper)}
were chosen with the aim of obtaining values of \postrefereechanges{$m$}
that are logarithmically close to the number of administrative
divisions in the Iranian 2009 presidential-election first round.  The
alternative to choosing the ``electoral zone'' level in the Brazilian
election would have been to choose the state/federal district level,
giving \postrefereechanges{$m=27$.}  
The Iranian 2005 election first round should be that
which most closely mimics the statistical characteristics of that of
Iran in 2009. 
\postrefereechanges{Criterion (5)} does not necessarily imply that the
elections were totally free of artificial interference. However, if
some artificial interference did occur in these control elections,
then that should most likely lead to an underestimate of the
significance of an unusual statistical signal in the Iranian 2009
presidential-election first round. Hence, this approach is 
conservative.

\tabfiginserted{\tpreselections}

\subsection{Pre-election polls} 
\label{s-prepolls-data}

At least twenty-seven pre-election opinion polls of Iranians' voting
intentions were carried out from March to June 2009 and were
publicly known and referenced by the English-speaking community
before the day of the election \citep{Gerash772009a}. 
Among these,  %% 27  
six concerned Tehran only, %% 21   
two only gaving rankings rather than percentage support, %% 19
one concerned impopularity rather than popularity,  %% 18
and one only concerned employees, %% 17
leaving 17 nationwide polls from 11 journalistic sources.

The sources of these polls are diverse. 
Several polls are reported by news organisations or from sources
that are generally considered to be close to one or more of the
candidates. Thus, systematic biases could be suspected to be present either 
favouring the incumbent candidate or favouring candidates
representing either reformist or conservative opposition sectors. 
The selection process implicit in the creation of this list is a 
stochastic process that depends on the various positive and
negative feedback mechanisms in the editing of the English-language
Wikipedia in general, the way that these processes applied to the
English-language Wikipedia article concerning the Iranian 2009
presidential election in particular, and the way that English-Persian bilingual
editors participated in adding to the 
English-language article and correcting errors,
possibly improving consistency between the versions of the page in the two
languages.

From the article's creation on 2 May 2007 to the last edition on 
the day (UTC) preceding the election, 11 June 2009, 
a total of 345 edits were made
\citep{Gerash772009b}. The number of individual human authors
(excluding robots and grouping together authors identified by nearly
identical Internet Protocol numbers) is about 85. From the edit made
on 10 April 2009 adding information on an opinion poll up to 11 June
2009, a total of 212 edits were made.  Editing activity on the
corresponding Persian-language article started later and was more
intense. From its creation on 25 August 2008 to the last edit on 11
June 2009 (UTC), a total of 969 edits were made \citep{Agha2009}. The
number of individual human authors (excluding robots and grouping
together authors identified by nearly identical Internet Protocol
numbers) is almost identical to that of the English-language article,
i.e. about 84.  Information on pre-election polls in the
Persian-language article is not in the same tabular format as in the
English-language article. Consistency between different language
versions of a Wikipedia article is decided by consensus among
bilingual editors. About seven authors have edited both language
versions.  

These large numbers of editors and edits on the English and
Persian versions of the Wikipedia article, together with the highly
open and well-documented nature of the editing process, 
suggest that the
English-language compilation of pre-election opinion polls 
of the evening of 11 June 2009 is probably
less biased than any other compilation available 
prior to the date of the election.
Table~\ref{t-prepolls} lists these polls.\footnote{Another pre-election poll, carried out by non-Iranian organisations 
by telephone in early May 2009,
became known to the English-language online community {\em after} the election \protect\citep{TFT09}. Inclusion of this additional
poll would not modify the results significantly.}

\tabfiginserted{\tprepolls}

\subsubsection{Pre-election poll analysis} \label{s-prepolls-meth}

While the list of pre-election polls given in Table~\ref{t-prepolls}
is probably less biased than any other English-language compilation
available prior to the date of the election, at least two reasons can
be invoked to exclude some of the 17 polls listed in
Table~\ref{t-prepolls}. The two subsets of polls which could be excluded generally disfavour
candidate A and favour candidate M.

Firstly, the sources of seven polls (favouring candidate M since May 2009 and
candidate R five days before the election) are stated as being a combination
of a field survey and an internet survey, with a warning regarding a
likely demographic bias in favour of voters connected to the internet
\citep{tabnak13880310,RahbordedaneshPolls0319}.
Moreover, robots can credibly introduce significant systematic errors
in an internet survey, but (even as of 2012), are extremely unlikely to introduce
any errors in the raw datasets of telephone or door-to-door surveys. 
Unless additional information estimating the systematic errors in these
surveys is added to the analysis, 
the exclusion of these polls needs to be considered.

Secondly, two sources (favouring candidate M) are no longer online
\citep{GhalamnewsPolls0306,BaznevisPolls0316}.
No online archival copy is known, so 
their data remain public only indirectly \citep{Gerash772009a}.
Hence, exclusion of these two polls should be considered. 
Alternatively, censorship probabilities could included in the
analysis. Websites more likely to favour candidates disfavoured by the government
could be considered to have a higher chance of being closed down by administrative,
political or financial methods. In order to derive results that are conservative
in the sense of favouring pro-government candidates, these probabilities are neglected
in this paper.
%Alternatively, the addition to the analysis of information
%regarding the suppression of websites reporting polls that favour
%one or more candidates could be used to 
%favour inclusion of these two polls in order to reduce overall bias
%in the set of polls. Such information is not used here.

Elementary statistical tests are applied to different subsets
of the pre-election poll data shown in Table~\ref{t-prepolls},
excluding either 
the partially internet-based polls, 
the polls that do not have a known online archive, 
or both, to see if the various subsets are self-consistent, and if they 
are consistent with the official results \citep{MOI09}.
Moreover, after applying both exclusions, it still remains credible that
among the remaining eight polls, 
either those polls favouring 
candidate A or those disfavouring A could contain
systematic errors favouring or disfavouring A, respectively. 
Thus, both of these subsets are considered separately, for self-consistency
and for consistency with the official results.

\tabfiginserted{\fsigabs}

\tabfiginserted{\fsigrel}

Can the stated numbers of people interviewed be used to assume that
the error in a poll is dominated by the random error implied by the
Poisson limit associated with the number of people interviewed, i.e.
assuming negligible additional error from the 
required demographic profile corrections? This would seem unrealistic.
For example, the prior-to-7 June IRIB survey
\citep{AlefPolls0318} and the 26 May--5 June ILNA survey \citep{Ilna},
taken during approximately the same period, 
should differ in their results by less than about 3\% (95\%
confidence) if only the Poisson limit errors are considered.
Yet their estimates for candidates A and M differ by
36.5\% and 27.3\%, respectively.  Clearly, the errors in at least one of
these two cases are about an order of magnitude larger than the
minimum random error attainable for the stated sample sizes. 
Moreover, nine of the polls do not state their sample size at all.

\tabfiginserted{
  \fhistall

  \fhistA
  \fhistR
  \fhistK
  \fhistM
}

Hence, for a conservative analysis of the data, linear least-squares
best fits of voting intentions as a function of civil date $t \in [1,100)$
in days, where $1=$ 5 March 2009 and $100=$ 12 June 2009 are made
for the different subsets, giving equal weights to all polls in a given
subset. The quality of the fit is estimated using $\chi^2$
of the residuals, where the variance $\mathrm{Var} (w_{ij}) = (3\%)^2$ is assumed. 
This correponds to assuming that the error is no more than the random error for
a survey of 1000 people obtainable from the Poisson limit. 
The cumulative probability distribution of the $\chi^2$ distribution 
then gives an estimate of the degree to which the polls of a given subset
agree with each other.

Since the assumed variance should in principle be an overestimate, but
in practice may be an underestimate, the standard errors 
from the fits are then used to estimate the uncertainty in the expected results 
of the 12 June vote for the majority candidates A and M.
The errors are assumed here to be random and normally distributed.
This is clearly only a rough approximation. Linear fits to the complement
of the major candidates' pre-election support are also considered.

\tabfiginserted{
  \fcalib
  \fbooteg
}

\section{Results} \label{s-results}

Figure~\ref{f-sigabs}  shows that the two
Iranian presidential elections have a higher dispersion in the 
total population sizes of voting areas (shahrestans) than the elections
used for comparison. Unsurprisingly, the dispersions in the 
logarithmic vote counts for the individual candidates are also higher
in the Iranian elections compared to the other four. Thus,
the Iranian elections are more likely to approach the
Benford's Law limiting case for first digit distributions than the
other elections, since their logarithmic distributions are wider.
Moreover, the widths are roughly $0.5$, so that, to the degree that
each distribution of $v_{ij}$ is log-normal, about 68\% of the 
corresponding cumulative distribution should cover one decade. This is another
factor in favour of the Iranian elections approaching the Benford's
Law limit.

Figure~\ref{f-sigrel} 
%Figure~\ref{f-sigabs} (right)
shows that this is not just an effect of the
dispersion in the population sizes of the voting areas. The voting
rates $\widetilde{v_{ij}}$ in the Iranian elections generally are
more widely log-dispersed than in the other four elections. 
Figure~\ref{f-histall} shows the distribution of the total votes
per voting area, and 
Figs~\ref{f-histA}--\ref{f-histM} show 
%Fig.~\ref{f-histA} shows
the distributions of the votes for the four candidates.

\subsection{Empirical, local bootstrap model} \label{s-calib-res}

Figure~\ref{f-booteg} 
%Figure~\ref{f-calib} (right)
shows an example local bootstrap, 
as described in Definition~\ref{d-local-boot},
for an arbitrarily selected candidate (the first in the table) 
in an arbitrarily selected election (the French 2007 election
first-round).
This illustrates how closely the local bootstrap imitates the 
data set. The logarithmic dispersion of voting rates $\widetilde{v_{ij}}$ 
for this candidate is greatest in highly populated voting areas, i.e.
for $x_i \sim 10^6$. The bootstrap imitates this wide dispersion
at $x_i \sim 10^6$ with little obvious effect on low population voting areas.
At lower populations, there are some outliers with low voting rates in the real data. 
The bootstrap realisation appears to reproduce similar outliers, while also
matching the bulk of the $\widetilde{v_{ij}}$ distributions in these lower
population voting areas. Overall, the realisation appears to closely
match the distribution of the official data, possibly exaggerating the
numbers of outliers. Subjectively, the figure suggests, as expected, that
the simulations should be conservative. They are likely to statistically
mimic features of the data even if those features constitute artificial
interference in the data set. Only a very unusual interference in the
data set is likely to be considered extreme when compared to an ensemble
of local bootstrap simulations.

Another characteristic that is clear in the figure is that which 
is intrinsic to bootstrap methods: a particular value of $\widetilde{v_{ij}}$
may occur several times
among the values $\viprimejhat$, introducing a discreteness effect.
However, in the present method, what is of interest for first digit frequencies
is $\viprimejhat x_i$. This should remove most of this discreteness effect.

In order to estimate bootstrap confidence levels ${\cboot}_j(d)$ 
%% [Eq.~(\ref{e-defn-cbootjd})] 
for rejecting each of the candidates'
first digit frequencies in each of the elections in the control data set, 
$10^5$ local bootstrap simulations were generated.
Together with 
$\cempir$ defined by the actual frequency distribution of these values,
least-squares fits were found numerically as described above, giving
\begin{equation}
% bm =  1.824142 bp =  1.487156 cbs =  0.5660550
\beta^- = 
  1.824
,\, \beta^+ = 
   1.487
,\, \cbootstar = 
    0.566
\label{e-calib-solution}
\end{equation}
in Eq.~(\ref{e-cempir-cboot}).

\tabfiginserted{
  \fAconflev
  \fRconflev
  \fKconflev
  \fMconflev
}

Figure~\ref{f-calib} shows that the local bootstrap confidence
intervals are biased and skewed. They are 
conservative at high confidence levels at both the lower and upper limits. 
Taken literally, 
the local bootstrap confidence
intervals fail to reject the actual first-digit 
frequencies as often as they should. 
For example, none of the 459 first-digit frequencies for 
candidates in the control elections
fall below the lower cutoff of the 95\% confidence interval, i.e.
the 2.5\% percentile, as 
calculated directly by the bootstrap method for each of those candidates,
even though in principle about 11 should. %% 0.025 * 459 = 11.475
The calibration given 
by Eqs~(\ref{e-calib-solution}) and
(\ref{e-cempir-cboot}) and illustrated in Fig.~\ref{f-calib} should provide
more accurate (less conservative) estimates of confidence levels.

\tabfiginserted{\fprepolls}

The inverse of Eq.~(\ref{e-cempir-cboot}) can also be useful, i.e.
\begin{equation}
\cboot(\cempir) = 
\left\{ 
\begin{array}{l c c}
   \left( {2 \cempir} \right)^{1/\beta^-} \, \cbootstar
  & , & \cboot < \cbootstar \\
  1 -  \left\{ 
  \left[ 2 \left({1-\cempir} \right)\right]^{1/\beta^+} 
  \right\} \left( 1- \cbootstar \right)
  & , & \cboot \ge \cbootstar,
\end{array}
\right.
\label{e-cboot-cempir}
\end{equation}
using $\beta^-, \beta^+$ and $\cbootstar$ from Eq.~(\ref{e-calib-solution}).

\subsection{Iranian 2009 presidential election} \label{s-BL-res}

Figures~\ref{f-Aconflev}--\ref{f-Mconflev} show 
%Figure~\ref{f-Aconflev} shows
the first digit
frequencies of the four candidates of the Iranian 2009 
first-round presidential election using the corrected confidence
intervals $\cempir$ corresponding to two-sided confidence levels
of $95\%, 99\%$ and $99.9\%$.

It is clear in 
Fig.~\ref{f-Kconflev} 
%Fig.~\ref{f-Aconflev} (bottom-left)
that candidate K has an excess
frequency of votes $v_{ij}$ that start with the digit 7, to a 
calibrated significance well above $99.9\%$ (two-sided). 
The bootstrap
estimate is ${\cboot}_{\mathrm K7} > 99.924\%$, and the calibrated estimate is
${\cempir}_{\mathrm K7} > 99.9960\%$, i.e. 
$1- {\cempir}_{\mathrm K7} \approx 4 \times 10^{-5}$. Since this could be considered
one of 36 independent tests for the full Iranian 2009 data set, 
a \v{S}id\`ak-Bonferonni correction factor \citep{Abdi07} of 36 
gives
\begin{equation}
  \pseven < 1.5 \times 10^{-3}
  \label{e-pk7-sidakbon}
\end{equation}
for rejecting the full data set.
Even if the bias/skewness correction were ignored, i.e. if $\cboot$ were used,
the full data set would still be rejected at a level of $p < 3\%$ based on the 
excess of first-digit 7 votes for K, i.e. 
$36 (1- {\cboot}_{\mathrm K7}) \approx 0.027$.

\subsection{Pre-election opinion polls}
\label{s-prepolls-res} 

\subsubsection{Candidates A and M}

\tabfiginserted{\tprepollsprob}

\tabfiginserted{\tprepollsbest}

Figure~\ref{f-prepolls} shows the pre-election poll data 
(Section~\ref{s-prepolls-data}, Table~\ref{t-prepolls}).
Both panels include equal-weighted, 
linear, least-squares fits to the temporal evolution of the candidates' 
support, which are extrapolated to the election date. 
Interpreted literally, the extrapolations of the linear fits to 
the election data would imply that candidates A and M each won about 40\% of 
the votes, i.e. a second-round of the election was necessary.
However, it is also clear that this is a best-fit that differs from
the individual poll results for A and M 
by $\sim$10--20\% during the fortnight preceding
the election. This is the case for both panels, i.e. whether including or
excluding the partially internet-based and publicly unarchived polls.

For candidates A and M, the quality of 
the linear fits in these two panels,
assuming $\pm 3\%$ random, normally distributed errors (68\% confidence) as stated above
(Section~\ref{s-prepolls-meth}), is given 
for the full set of polls 
by   
$\chi^2\approx 305$ and $194$, respectively, for $d=16$ degrees of
freedom, and for the smaller set of polls by 
$\chi^2\approx 188$ and $92$, respectively, for $d=7$.
This is why the confidence levels, shown in 
the first and fourth rows of Table~\ref{t-prepolls-prob}, respectively, show the
$\chi^2$ confidence levels estimated as $p_{\chi^2} \approx 0$.

Table~\ref{t-prepolls-best} shows the estimates of what the results of
the first-round election on 12 June should have been if the linear
fits and the residuals to the fits are used. Even in the fourth row,
where the exclusions have been applied and five out of the eight remaining 
polls are from sources that 
could be suspected of being more likely to favour candidate A than candidate
M \citep{presstv2pollsJune1,presstv2pollsMay12,AlefPolls0318}, the best
estimate is that no candidate was expected to obtain over 50\% support on 12 June.
%% [7.4 5.2] ./ ((1 ./ sqrt( sum( [ ir2009a(:,3) ir2009a(:,6) ]))) * 100)
The precision is about 200--400 
times worse than that of the official results, where
a Poisson error would be about $0.02$--$0.03\%$ for 
candidates A and M, rather than about 5--7\% as estimated here.
This is partly because of the disagreement between the different polls.
Unless an assumption about which subset of polls is the most reliable
is made, it is difficult to make preciser estimates. 
Hence, assuming normal error distributions, 
the probability of a first-round win by candidate A or M is about
$p < 23\%$ or $p < 26\%$, respectively, from the fourth row of 
Table~\ref{t-prepolls-best}. This is not a significant rejection of a win
by either candidate. A larger subset of the polls (higher row in the table) 
can increase or decrease the probability of a first-round win by M, 
but in all three cases gives a significant rejection of a first-round win
by A. If the official results have been interfered with, 
then the estimate in the fourth
row of Table~\ref{t-prepolls-best} may be more {\em accurate} than the official
results, despite being a few orders of magnitude less {\em precise}.

The final two rows of Table~\ref{t-prepolls-best} show that if either
the five polls favouring or 
the three polls 
disfavouring candidate A (after the
exclusions have been applied) are considered alone, then either
candidate A or M is expected to have 
very significantly or insignificantly 
won the first-round election, respectively, assuming normal error
distributions. The corresponding 
values of $p_{\chi^2}$ in the last two rows of Table~\ref{t-prepolls-prob}
show that the self-consistency among the polls in either subset 
considered alone is quite
high. The worst case is $p_{\chi^2}(\mathrm{M}) = 0.050$ for polls disfavouring
A. Inspection of the lower panel of Fig.~\ref{f-prepolls} suggests that
this is because $w_{i{\mathrm M}}(t_i)$ has a negative second derivative 
in this case, 
i.e. M's support appears to have saturated at a little over 50\% about a week
before the first-round election, if this subset of polls is considered
alone and without the addition of systematic error.
These results are unsurprising: polls favouring A are self-consistent and
favour A, and polls disfavouring A are marginally self-consistent for a linear
fit and favour M.

\subsubsection{Candidate K}

While conclusions regarding whether candidate A or M or neither
won the election first-round are weak due to the 
disagreement between the pre-election polls, 
the upper panel in Fig.~\ref{f-prepolls} and the first and third
rows of Table~\ref{t-prepolls-prob} show that despite this disagreement,
all the polls that published estimates of K's support
agree with each other for the best linear fit for a $\pm 3\%$ assumed error.
Moreover, the standard error in the fit on the day of the 
election first-round is small. The first and third rows of 
Table~\ref{t-prepolls-best} show that for the two subsets that contain
sufficient information on K's vote support in order to make a linear
fit to the data, the expected result for K is 
$w_{i{\mathrm K}}^* \approx 7\pm 1\%$, with a negligible dependence on 
inclusion/exclusion of the two publicly unarchived polls.

This very strongly disagrees with the official result in the 
MOI data \citep{MOI09}. 
Candidate K only got 0.838\% support in the official result.
This is about six standard errors lower than
the mean. This is why $p_{y_{\mathrm K}} \ll 1 $ in Table~\ref{t-prepolls-prob}.
The slope of the fit is $(-1\pm 3) \times 10^{-4}$, i.e. 
$(-1 \pm 3)\%$ over 100~days, for both poll subsets. That is, 
K's support is consistent with being constant over three months.
Assuming that K's support did not change at all gives only slightly
differing results to the above: 
$w_{i{\mathrm K}}^* \approx 7.4\pm 1\%$ and $p_{y_{\mathrm K}} \sim 10^{-15}, 10^{-12}$ for 
the two poll subsets.

In other words, what the pre-election polls best agree on is
that K's support was very close to constant for approximately 
three months. They agree on this with a margin of error that
is reasonable for carefully carried out opinion polls 
of typical sample size ($\sim 1000$) and 
consistent with the scatter among the polls' estimates for
K's support. These same polls reject the official result
to very high significance. 

Nevertheless, most of the polls that reported on support
for candidate K are partially internet-based.
Hence, several interpretations of this result are possible, including:
%\begin{list}{(\roman{enumi})}{\usecounter{enumi}}
\begin{list}{(\arabic{enumi})}{\usecounter{enumi}}
\item 
  there was a swing against K during the last several days preceding the
  election, i.e. he lost about 90\% of his support during the days
  preceding the election; or
\item 
  the partially internet-based polls overestimated K's support by
  about a factor of 10; or
\item 
  the official result for K was artificially modified.
\end{list}
Case (1) seems unlikely. After three months of
stable support, it seems sociologically unrealistic that 90\% of
K's support would disappear in several days, barring a
dramatic, well-mediatised event that motivated
an overwhelming majority of previously ``committed'' supporters
of K to suddenly drop their support. 
The highly significant excess of
vote counts with first-digit ``7'' for candidate K found using the
local bootstrap method (Section~\ref{s-BL-res}) is consistent
with case (3).
However, 
without a lot more information on the calibration methods and demographic
profiling used in these polls, it would be difficult to exclude
case (2).

\subsubsection{Candidates R, K, and ``other'' (O)}  \label{s-prepolls-res-RKO}

Could it be possible to estimate the expected 12 June result for K
without relying on the partially internet-based polls, despite
the fact that most other polls did not state the suport for K?
Contingent on some credible assumptions, yes, though indirectly.
The complement of 
the polls' estimates of support for candidates A and M can be used.
In principle,
this should represent the sum of support for candidates R and K and 
interviewees that were undecided or gave ``other'' responses to
the interview question. 
This combined voter component 
is designated  as ``RKO'' in Fig.~\ref{f-prepolls} 
and Tables~\ref{t-prepolls-prob} and \ref{t-prepolls-best}.

In fact, Table~\ref{t-prepolls} shows that in those polls including information
for all four candidates, the sum of the four candidates' support,
or that of the five candidates in the polls of 5 March, 4
    April, and 5 May [where support for 
candidate Kh, who announced his withdrawal on 16
March, is estimated at 36.2\%, 27.6\%, and 5.3\%, respectively
\citep{RahbordedaneshPolls0319}],
are 
$100\%\pm1\%$, except for the prior-to-6 June poll, where 
the sum is 98\% \citep{BaznevisPolls0316}. 
Since the fractions of undecided/other voters are rarely this small,
especially weeks or months prior to an election, 
this suggests that the tradition
in the reporting of opinion polls in Iran is to normalise away undecided/other
responses. 
If this speculation is correct, 
in the sense that all the polls listed follow this tradition,
then 
\begin{equation}
  w_{i{\mathrm{RKO}}} := 1 - w_{i{\mathrm{A}}} - w_{i{\mathrm{M}}}
\end{equation}
should be a good approximation to $w_{i{\mathrm{RK}}}$, i.e. the sum
of the support for the two minor candidates, without the complication 
of undecided/other voters. However, since this is only a speculation,
the designation RKO will be retained initially.

\tabfiginserted{\tprepollsbestlinear}

\tabfiginserted{\tprepollsbestquadratic}

\tabfiginserted{\tprepollsbestcubic}

The lower panel of Figure~\ref{f-prepolls} and the RKO columns
in Table~\ref{t-prepolls-prob} show that just as the larger subset
of polls best agree internally on candidate K's vote support and 
strongly disagree with the official result for K, 
the situation for RKO is similar. 
As the second, fourth and lower rows of Table~\ref{t-prepolls-prob} 
show, $p_{\chi^2}$ is higher for RKO than for A and M in any given
poll subset, i.e. the vote support estimates for RKO 
are more self-consistent among the polls for RKO rather than for A or M. 
The linear best fits for RKO also 
strongly reject the official result for RK if the ``other'' component
is neglected. These statements hold no matter which subset
in Table~\ref{t-prepolls-prob} is chosen, even if the subset of five
polls favouring A is analysed alone. 
The rightmost two columns in Table~\ref{t-prepolls-best} show the
expected result $w_{i\mathrm{RKO}}^*$ for the different poll subsets.
It is clear why the official result
$y_{\mathrm{RK}} = 2.5\%$ is rejected to high significance.
However, these rejections depend on the role of the ``other'' component.
Possible interpretations include:
%\begin{list}{(\roman{enumi})}{\usecounter{enumi}}
\begin{list}{(\arabic{enumi})}{\usecounter{enumi}}
\item 
  some or most of the polls in any subset are not normalised polls, so that
  $w_{i{\mathrm{RKO}}}$ includes a component of $\sim 5$--15\%  of undecided
  voters, and most of this category of voters made a decision during the last
  several days to vote for A or M, or did not vote; or
\item 
  most or all of the polls are normalised polls, and there was a
  swing against R and/or K during the last day or two preceding the
  election, i.e. they together lost about 70--90\% of their support during
  the last several days preceding the election; or
\item 
  the official result for R and/or K was artificially modified.
\end{list}

Hence, similar results are obtained whether the partially internet-based
polls are included and K's pre-election support and official result are
analysed, or if the partially internet-based polls are excluded,
the remaining polls are assumed to be normalised, and the combined
vote RKO (equal to RK by assumption) is analysed instead. The possibility
of ``other'' voters in pre-election polls becoming invalid voters in
the official poll has little effect on the above three possibilities,
since the percentage of invalid votes was officially only about 1\%.

\subsubsection{Sensitivity to linear assumption} \label{s-quadratic}

\postrefereechanges{Given the inconsistency of the pre-election poll
  data, it would be difficult to justify fitting a more complex model
  of the evolution of voters' intentions. Nevertheless, it would be
  interesting to see how sensitive the conclusions are to the
  assumption of a linear relation. Given that fractional voting
  intentions $w_{ij}$ are bound between zero and unity, and that their
  sum is unity by definition, it is not trivial to choose the most
  realistic non-linear function for fitting the data. Linear fits
  themselves are not bound by the unit interval, as can be seen by the
  extrapolated negative support for candidate R in
  Fig.~\ref{f-prepolls} (upper) in early March 2009.}

\postrefereechanges{In order to qualitatively examine at least some
  alternatives to a linear fit, quadratic and cubic least-squares
  equal-weighted best fits have been carried out, using 10,000
  numerical realisations in each case in order to simplify calculation
  of standard errors.  In comparison to the method in
  Sect.~\ref{s-prepolls-meth}, this procedure makes it easier to make
  the more accurate, but still conservative, assumption that each
  survey concerned 1000 valid interviewees, and calculate Poisson
  errors per candidate, rather than assume a 3\% error in all cases.
  This is still conservative in the sense that all surveys with
  published numbers claim higher sample sizes.  Each realisation
  generates a simulated set of poll results by starting with each
  known poll result, calculating the Poisson error for the number of
  interviewees favouring a given candidate, and offsetting the known
  result by a number drawn from a normal distribution with this
  standard deviation.  A quadratic or cubic least-squares fit is made
  to each realisation. An ensemble of realisations (with either
  quadratic or cubic fits) is used to estimate standard errors.  Since
  this method differs from that used above, and should yield smaller
  standard errors, a linear fit was also performed with this method, in 
  order to enable direct comparison.}

\postrefereechanges{Tables~\ref{t-prepolls-best-linear}--\ref{t-prepolls-best-cubic}
  show the equivalent expected values and standard errors for
  the first-round election result when linear, quadratic, and
  cubic least-squares fits are found numerically this way.
  The results are not generally sensitive to the degree of the
  polynomial chosen for fitting. The strongest result from the pre-election
  poll data, that the expected RKO support is much greater than 
  the official result of $y_{\mathrm{RK}} = 2.5\%$, appears to be
  robust. A few cases are sensitive to the linearity assumption, 
  especially for candidates A and M when subsets of polls are analysed.
  For example, the fourth row of 
  Tables~\ref{t-prepolls-best-quadratic} and \ref{t-prepolls-best-cubic}
  shows that 
  a first-round win by candidate A is likely in both the quadratic
  and cubic cases, provided that both the partially internet-based
  polls and the publicly unarchived polls are excluded. However,
  to get close to the official result of $y_{\mathrm{A}} = 62.5\%$, 
  the polls disfavouring candidate A also have to be excluded
  (fifth row).}

\section{Discussion} \label{s-disc}

The excess of first-digit 7 votes for candidate K seen in
%Fig.~\ref{f-Aconflev} 
Fig.~\ref{f-Kconflev} 
implies a rejection of the null hypothesis,
i.e. the hypothesis that no artificial interference occurred in the
first-round Iranian 2009 presidential election results, at a
confidence level for the full data set of $\pseven < 1.5 \times 10^{-3}$
[Eq.~(\ref{e-pk7-sidakbon})].  This estimate is obtained by a highly
conservative method based on the data themselves---a local bootstrap
method---and calibrated using similar, prior elections. 
In other words, it is very difficult to explain this excess of
``K7'' votes by assuming that among voting areas of approximately the
same voting population size, the actual votes for a candidate can be
modelled as being randomly drawn from the statistical pattern
(distribution) of voting rates for voting areas of about that size for
that candidate. 
Patterns as unusual as
the excess of 7-something votes for K did not occur in 
the five previous presidential elections studied in the present work. 

Are there caveats that could weaken this result? What alternative
possible explanations, including artificial interference, may there be
for this excess?

\subsection{Small numbers of data values}

The actual number of voting areas whose vote counts for K start with
the digit 7 (``K7''{}'s) is 41. The Benford's Law limiting expectation
for this digit, for 366 voting areas, is 21.2, and the upper
confidence limits shown in 
%Fig.~\ref{f-Aconflev} (bottom-left)
Fig.~\ref{f-Kconflev}
lie in between
these two values. Intuitively, it may seem difficult to believe that just a few
dozen or so voting areas with 7-something votes are unusual enough to
reject the credibility of the whole data set. The intuitive error here
is probably related to not thinking through the process that (assuming
no interference in the data) generated the data. Large numbers of
Iranians thought individually, communicated with each other in
networks and groups, and eventually made some marks on paper
indicating their voting decisions. If Iranian voters in voting areas
of about the same size, voting for a given candidate, can be validly
modelled by a statistical distribution of a shape determined by the
official election results, then there is not much numerical freedom
for the vote numbers to vary other than according to 
that distribution.

On the contrary, the smallness of the number of unusual votes 
could be considered a factor in favour of 
an alternative hypothesis of artificial interference, if artificial
interference is to be made in a way that is perceived to be unlikely
to be detected. Someone wishing to 
interfere in the data might underestimate the natural constraints
for self-consistency within the data set. That is, s/he 
might intuitively, but incorrectly, guess that modifying a dozen 
or so values in a table of over a thousand values
is statistically insignificant, and, thus, undetectable.

\tabfiginserted{\fhistKfolded}

\subsection{Copying error}

Could the excess of K7's just be a copying error by employees under
pressure in a stressful situation? Various sources of unintentional
(but artificial) errors are possible. The present analysis only
concerns the data as published by the MOI, not a version of the
data closer to the ``raw'' values provided by individual voting centres.
However, this number remained stable from the first to third 
versions of the data.
The number of entries
that start with 7 under candidate K in the Arabic-Indic numerals
{\sc pdf} file is 41 \citep{MOI09c}, in agreement with that of
the first {\sc xls} file \citep{MOI09}.

\subsection{Could K's vote distribution be especially spiky/noisy?}

Logarithmic intervals of first digits give unequal weights to the
different digits if an intrinsic distribution is logarithmically
uniform. Could K's vote distribution be in general spiky (noisy), 
for a reason unrelated to human perception of the decimal system,
with some spikes hidden in the low digits (1, 2) because they 
correspond to wider intervals? Figure~\ref{f-histKfolded} suggests
that this is not the case. The excess of vote counts starting 
with 7 is the only obvious spike in K's vote distribution for
bins of a size and offset that match the first-digit 7 bin.

Moreover, another property of K's vote distribution suggests that 
this should be {\em more} smooth than typical vote distributions.
The frequency distribution of votes $v_{ij}$ is a convolution of
the distributions of $v_{ij}/x_i$ together with the distribution of
effective population sizes $x_i$ of the voting areas. 
The narrower the distributions $v_{ij}/x_i$ are, the less the convolution
will smooth out the underlying distribution of $x_i$, and in turn, the
less the first-digit frequency will be smooth. 

\tabfiginserted{
  \fKsevenA
  \fKsevenR
  \fKsevenK
  \fKsevenM
}

In 
%Fig.~\ref{f-sigabs}, 
Fig.~\ref{f-sigrel}, 
the second highest value of $\sigma[\log_{10}
  (v_{ij}/x_i)]$ of any of the presidential elections is that of
candidate K. The four widths $\sigma[\log_{10} (v_{ij}/x_i)]$ of the
votes for A, R, K, and M are 0.12, 0.34, 0.42, and 0.24,
respectively. Candidate K is the candidate in the first-round 2009
Iranian election who should {\em least} be expected to have a noisy
first-digit frequency distribution. The 68\% width (assuming
normality) of the logarithmic $v_{i\mathrm{K}}/x_i$ distribution is
approximately an order of magnitude.  So unless the shape of the
$v_{i\mathrm{K}}/x_i$ distribution is exceptionally spiky and
correlated between different voting population sizes $x_i$, the
resulting first-digit frequency should be very well smoothed out. The
narrowest width is for candidate A. 
Figures~\ref{f-Aconflev}--\ref{f-Mconflev} do 
%Figure~\ref{f-Aconflev} does
suggest that
apart from the excess of first-digit 7 votes (and slight lack of
first-digit 6's and 8's) for K, A's first-digit frequency distribution
is the most spiky of the four candidates' distributions, even when
the local bootstrap method is used to estimate confidence levels.

Hence, spikiness in K's vote distribution seems to occur
uniquely in 
%(and immediately adjacent to)
the first-digit 7 bin, and it does so in a way that manages
to avoid a smoothing effect that is stronger than for the other candidates
in the same election and nearly all candidates in the five control
elections.

\subsection{Does an interference hypothesis lead to any further unusual
properties?}

\subsubsection{The six most populous voting areas and the excess $7d_2d_3d_4$ vote counts for K}

If the excess 7's for K indicate
interference in the data, then other signs of interference could be
expected. A reasonable hypothesis would be that vote counts
for one of the major candidates were decreased or increased.
Figures~\ref{f-KsevenA}, \ref{f-KsevenR}, \ref{f-KsevenK} and \ref{f-KsevenM} show
%Figure~\ref{f-KsevenA} shows 
proportions of votes that each candidate received as a function
of total votes, where those voting areas selected by having
7 as the first digit in K's vote count are highlighted.

\tabfiginserted{\tKseven}

{\em It is clear in Fig.~\ref{f-KsevenA} that for candidate A, among the
  six voting areas with the greatest numbers of total votes, the three
  of these that voted for A in the highest proportions are all
  selected by the K first digit 7.} 
Data for these six voting areas are listed in Table~\ref{t-Kseven}.
Fig.~\ref{f-KsevenM} 
shows correspondingly
that the three of the six most populous voting areas who voted least for
M are also those selected by the K first digit 7. Since the
vote fractions for K and R are 
(in the MOI data)
only about 1\% each, the high proportions of 
votes for A necessarily imply low proportions of votes for M.

Are these two subsets of the set of the six most populous voting areas, 
the $d_1(\mathrm{K})=7$
group voting over 60\% for A, versus the $d_1(\mathrm{K}) \not=7$
group voting less than 55\% for A, significantly distinct? 
The number of points is very small, but the separation between
the two populations seems to be clear.
The Kolmogorov-Smirnov (KS) test is a 
{non-parametric} test that enables the comparison of these
two populations.  
Among the six voting areas with the highest vote numbers,
the probability that the three vote proportions for A
$\{ v_{i\mathrm A}/x_i \;|\; d_1(\mathrm{K})=7 \}$ and the three vote proportions for A
$\{ v_{i\mathrm A}/x_i \;|\; d_1(\mathrm{K})\not=7 \}$ are sampled from
the same probability density function, 
or equivalently,  that the vote proportions
$\{ v_{i\mathrm A}/x_i \;|\; v_{i\mathrm A}/x_i > 0.60 \}$ and the vote proportions
$\{ v_{i\mathrm A}/x_i \;|\; v_{i\mathrm A}/x_i < 0.55 \}$ are sampled from 
the same probability density function, is 
\begin{equation}
\psevenabcKS \approx 0.100.
\label{e-psevenabcKS}
\end{equation} 
This high probability may seem counterintuitive, since the two sets of
three values are completely non-overlapping. However, the number
of values is extremely small. Non-parametric tests in general have less
statistical power than parametric tests, so it is unsurprising that the
test is weak. 

A stronger test can be applied if it is assumed that 
the two sub-samples can be approximated as being sampled from normal or 
log-normal distributions. In this case, 
the difference in the two means is about 3.7 or 3.5 standard errors in the mean,
respectively,
i.e. the probability that the two subsamples distinguished by the K excess
7's among the six biggest shahrestans have the same mean is rejected
at 
\begin{equation}
\psevenabcG < 5\times 10^{-4}.
\label{e-psevenabcGauss}
\end{equation}
%% $ 3\times 10^{-4}$ for normal distributions. 
%% $ 5\times 10^{-4}$ for log-normal distributions. 
%% => < 5 e-4.
The introduction of an assumption regarding possible distribution shapes
enables a much more significant rejection of the two samples being drawn
from a single distribution than the non-parametric approach.
The dependence
on the distribution shape (normal vs log-normal) 
does not appear to be strong.

Could it be expected that,
in the absence of artificial interference,
there is any statistical dependence between 
the initial first digit test leading to the excess of 7's for K
and the separation of the largest 
six voting areas into two distinct distributions using this
same characteristic? This seems unlikely.
%, so let $\psevenall$ and 
%$\psevenabcKS$ be considered as independent tests.

Another coincidence is obvious among the 
three $7d_2d_3d_4$ vote counts shown in Table~\ref{t-Kseven}: 
the second digit $d_2$ is zero in all three
cases. 
The standard form of 
Benford's Law should provide a reasonable approximation to 
the expected second digit distribution, especially given 
that K's vote distribution
is logarithmically wide, with $\sigma[\log_{10} (v_{ij}/x_i)] = 0.42$.
Benford's Law for the second digits 
gives the probability of 0 as a second digit to be 
11.97\%, i.e. slightly greater than 10\%. (The expected frequency 
of second digit $d$ is $\sum_{k=1}^{9} \log_{10} [1 + 1/(10k+d)]$.)
The probability that all 
three digits are identical---but not necessarily 0---is 
\begin{equation}
\psevenabczeros \approx 0.01037, 
\label{e-psevenabczeros}
\end{equation}
i.e. slightly
greater than $10 (0.1)^3$, which would be estimated assuming a uniform
{\em linear} distribution of values.

Again, this should be independent of the previous probability
estimates. There is no reason why dividing the six most populous
voting areas into two groups based on their first digit for K's votes
being 7, or being in the upper or lower half of voting proportions
for A, should have an effect on the second digits of K's votes.
%Together, the probability of accepting the null hypothesis is now
%% 0.00072 * 0.100 * 0.01037
%$\psevenall \;\psevenabcKS \;\psevenabczeros = 7 \times 10^{-6}.$
%

% sum( ([0.600 0.609 0.669] .- 0.5) .* [947168 1095399 1536106] )

If we suppose that (i) these three voting areas (Shiraz, Isfahan 
and Mashhad for 7078, 7002 and 7098 votes for K 
respectively)
should have proportions of about 50\% for A in agreement with
Tabriz and Karaj, which follow an approximately linear 
upper boundary to A's proportions of votes in the log-log plot
in Fig.~\ref{f-KsevenA}, and 
{if}
(ii) the total number of votes
should remain constant, then from Table~\ref{t-Kseven} 
this would imply that the correct number of votes for A 
would be about 473,000 less than in the MOI table. To keep 
the total number of votes constant, M's, K's and R's votes would
also have to be corrected. If these are corrected in proportion
to the three candidates' overall vote percentages, 
then the difference between A's and M's total
vote counts would be reduced by about one million votes.

\tabfiginserted{\tKseventies}

\subsubsection{\protect{Excess $7d_2$ vote counts for K}}

The $v_{i\mathrm K} = 7d_2d_3d_4$ voting areas comprise just
a small fraction of the total number of excess first digit 7 votes for K.
In Fig.~\ref{f-histK}, 
%Fig.~\ref{f-histA} (bottom-left),
a peak in the 70's, i.e. $7d_2$ 
votes appears to be strong. Table~\ref{t-Kseventies} shows these vote counts.

This distribution is quite literally odd. Most (15 out of 20) of the votes are odd, 
and the few even votes that occur are themselves distributed with perfect
uniformity. Every even number occurs exactly once.
The latter necessarily
implies the former---these two coincidences are dependent on one another.
Given that 
\postrefereechanges{20 randomly chosen integers lie} in the range from 70 to 79,
what is the probability for each even number to occur exactly once (and
by implication, for there to be a large majority of odd numbers)?
Alternatively, a more conservative estimate of how unusual this
distribution is would be to calculate the probability for at least 
15 out of the 20 integers in the range 70 to 79 to be all odd or 
all even. In both cases, the distribution among the 10 numbers should
be logarithmically uniform.

Numerical generation of 20-tuplets of numbers 
\begin{equation}
\{ \lfloor 10^{\log_{10}70 + x_k \log_{10}(8/7)} \rfloor \}_{k=1,20},
\end{equation}
where $x_k$ is selected uniformly on $[0,1)$, gives a sample of a
  logarithmically uniform distribution of numbers in the interval
  $[70,80)$.  This gives the probability that each even number occurs
    exactly once to be $p \approx 5\times 10^{-4}$.  
    The more conservative probability, i.e. 
    the  probability that either 
    at least 15 of the twenty $7d_2$ vote counts for K  are odd or
    at least 15 of the twenty $7d_2$ vote counts for K  are even,
    given that they are randomly
    selected from a logarithmically uniform distribution in the range
    $[70,80)$, 
    is (unsurprisingly) higher,
\begin{equation}
  \psevena \approx 0.04,
  \label{e-psevena}
\end{equation}
which is identical to the corresponding binomial probability to this
precision. Taken alone, this gives the odd dominance of 7$d_2$ votes for
K to be only marginally significant.

\tabfiginserted{\tKsevenall}

However, Table~\ref{t-K7-all} shows that the vote counts for the full set of 41 
K7-selected
shahrestans are dominated by odd votes for {\em all four candidates.} If 
the parity of the vote counts is an independent statistic, then
the overall probability is
\begin{equation}
  \pKsevenallodd \approx 0.00044,
  \label{e-k7-allodd}
\end{equation}
so the overall dominance of odd vote counts in K7-selected voting
areas {\em is} highly significant. 

%However, 
%it is again difficult to see how this characteristic could be
%statistically dependent on the initial selection based on the overall
%first digit statistics for the candidates' vote counts.

\subsection{Polling station observer crosschecks}

%% for discussion section

\tabfiginserted{\tpollobservers}

\citet{Brill10} states that all four candidates had thousands of
observers at individual polling stations
(Table~\ref{t-pollobservers}). He argues that since there were no
disputes about discrepancies between observers' records of individual
polling station results and the Ministry's statement of the same
numbers, there cannot be shahrestan-level interference in the vote
counts, since the latter must be the sums of the individual polling
station results, which were undisputed. Flaws that weaken this reasoning
include:
%\begin{list}{(\roman{enumi})}{\usecounter{enumi}}
\begin{list}{(\arabic{enumi})}{\usecounter{enumi}}
  \item The K7 anomaly firstly concerns candidate K.  Registered observers for K
    were absent from the vast majority of polling stations, even with only
    one observer per polling station (Table~\ref{t-pollobservers}).
  \item There was sufficient time for top-down artificial interference
    as hypothesised by \cite{BebScacco2009b}: large-scale results
    were released first, and the detailed results needed only minimal
    interference in order for the arithmetic to be consistent.  Tens
    of thousands of numbers can be summed in much less than a second
    on a 2009 personal computer.
  \item The number of M's observers is disputed between M and governmental
    authorities (Table~\ref{t-pollobservers}).
  \item Given the prepoll analysis here (Sect.~\ref{s-prepolls-res})
    and \citet{Meb09a}'s analysis, a credible hypothesis would be of
    major reductions in votes for R and K, major ($\sim 5$--$15\%$)
    additions of votes for A, and possibly minor reduction in votes
    for M. R and K had very few observers, so the chance of their
    observers discovering polling station discrepancies is
    low. Candidate A had many observers, but expecting them to
    complain about the Ministry overestimating the vote counts for A
    compared to what they observed is unrealistic.  People responsible
    for vote-count alterations were likely to have been aware of the
    relatively large number of M observers, and may have been aware of
    their localisation, in order to know at which polling stations
    interference would have been detectable. Even if the claim that M
    observers were present at 40,676 of the 45,692 polling stations is
    accepted, interference in the other 5,016 polling stations may
    have been sufficient to match the earlier province and shahrestan
    level results.
\end{list}

Nevertheless, \citet{Brill10}'s argument points to a method of
reducing the chance of interference in future Iranian presidential
elections.  Polling station crosschecks by observers for all
candidates at all polling stations, along with allowing mobile and
landline telephone networks \citep{Lucas09} and the internet to remain
active prior to, on and following the election day, would enable the
rapid, progressive release of the complete per-polling-station and
summed results by all candidates independently. Discrepancies could
then be traced and resolved individually.

\subsection{Post-election opinion polls}

Several opinion polls were carried out after the election first round
and revealed publicly more than six months later
\citep{Herman10}. Given the intense social conflict that occurred
following the election, a linear model of public support for the four
candidates across both the pre- and post-election periods is likely to
be a considerably less realistic model than that presented in the
opinion poll analysis in this work, which excludes the post-poll period. 
These post-election polls find
that shortly after and several months after the election first round, 
candidate A had about 55--65\% support. 

Accepting the post-election polls as accurate requires accepting the
sociological assumption that Iranians' opinions are accurately
described by statistical modelling.  If Iranians' voting preferences
can be modelled as random variables that depend mainly on demographic
profiles and are measurable using standard methods of random sampling,
then it is hard to reject the much more conservative local bootstrap
model, which uses the official vote counts {\em as their own model}.
In other words, accepting the post-election polls as statistically
meaningful requires accepting the official first-round election
results to be statistically anomalous. Statistical techniques cannot
be accepted in the case with a sample size of $N \sim 10^3$ and
rejected where $N \sim 10^{7.5}$ and fewer assumptions are required.
Thus, the statistical evidence for artificial interference in the
election results remains overwhelming, but the true results remain
uncertain.  One possible interpretation of the post-election polls is
that they could be seen as strengthening the case that if no
interference in the vote counts had occurred, then candidate A would
have won a first-round absolute majority (row 5 of
Tables~\ref{t-prepolls-prob} and \ref{t-prepolls-best}, excluding
partially internet-based polls, publicly unarchived polls, and polls
disfavouring A) and it would have been widely accepted as legitimate.
For opinion poll data to be more useful in the future, it would be
best if a variety of Iranian institutes, well-reputed for independence
and statistical methodology, carried out the polls and published
detailed reports of methodology and results.

\section{\protect\postrefereechanges{Summary and conclusion}} \label{s-conc}

A local bootstrap method was defined (Definition~\ref{d-local-boot})
in order to conservatively analyse vote-count first-digit frequencies
in presidential-election first rounds without assuming the Benford's
Law limit. Its validity is not necessarily restricted to this domain,
but the interest here is that the method can be applied to the 2009
Iranian presidential-election first round.  The method was calibrated
on a control set of five presidential-election first rounds
(2002--2006) and applied to the vote counts per voting area published
on 2009-06-14 by the Ministry of the Interior of the Islamic Republic
of Iran \citep{MOI09}.

The most deviant first-digit frequency is the excess of first-digit 7 votes
($d_1(v_{iK}) = 7$)
for candidate K. This is obvious in Fig.~\ref{f-Kconflev} and is also
present for uniformly spaced bins in the folded logarithmic vote-count
distribution in Fig.~\ref{f-histKfolded}. The uncorrected bootstrap
estimate of the significance of the excess implied by the 
control elections
is $1 - {\cboot}_{\mathrm K7} < 8 \times 10^{-4}$. 
The correction $\cempir(\cboot)$  
for bias and skewness derived from the control elections
is given in Eqs~(\ref{e-cempir-cboot}) and (\ref{e-calib-solution}).
The corrected significance is 
$1- {\cempir}_{\mathrm K7} \approx 4 \times 10^{-5}$. 
A \v{S}id\`ak-Bonferonni correction factor \citep{Abdi07} of 36 
gives the probability of finding a first-digit deviation this strong
in the full dataset of $\pseven < 1.5 \times 10^{-3}$ [Eq.~(\ref{e-pk7-sidakbon})].

This excess is all the more difficult to explain by any natural voting
process given that the logarithmic 
width (standard deviation) of candidate K's voting rate distribution 
$\sigma[\log_{10}(v_{i\mathrm K}/x_i)] = 0.42$ is the second greatest of 
that of any candidate among the Iranian 2009 and five earlier, similar datasets. This
width indicates a wide scatter in people's voting rates for any given
voting region size, which should smooth out the resulting vote counts
$v_{i\mathrm K}$ and push first-digit frequencies towards the Benford's Law
limit.

If the K7 anomaly is a genuine coincidence, i.e. if a chance numerical
coincidence that happens about once in a thousand times to statistically
equivalent datasets occurred rather than artificial interference in the
data, then selecting voting areas by this criterion should constitute
an arbitrary method of selecting a subset of the full data, and should
not lead to further unusual events.
However, K7 selection {\em does} lead to several coincidences, one
of which appears to 
favour candidate A, who officially won the election first round,
in the six most populous voting areas.
%\begin{list}{(\roman{enumi})}{\usecounter{enumi}}
\begin{list}{(\arabic{enumi})}{\usecounter{enumi}}
\item 
  Of the six voting areas with the greatest total numbers of voters,
  listed in Table~\ref{t-Kseven} (see also Fig.~\ref{f-KsevenA}),
  three of these (Shiraz, Isfahan, Mashhad) satisfy this criterion,
  i.e. they have vote totals for K that start with 7 (7078, 7002, 7098
  votes respectively).  All three of these have greater proportions of
  votes for A than the other three voting areas.  The probability for
  the two sub-groups of the six big cities to be drawn from the same
  distribution is mildly rejected by a non-parametric
  (Kolmogorov-Smirnov) test, with $\psevenabcKS \approx 0.100$
  [Eq.~(\ref{e-psevenabcKS})], but strongly rejected by a parametric
  (difference in means, assuming either normal or log-normal
  distributions) test, with $\psevenabcG < 5\times 10^{-4}$
  [Eq.~(\ref{e-psevenabcGauss})].
\item
  The other surprising property of the three big cities that have
  total votes starting with the digit 7 for K is that they all have
  the same second digit.  The probability for the second digit of all
  three to be equal (not necessarily zero) is $ \psevenabczeros
  \approx 0.01$ [Eq.~(\ref{e-psevenabczeros})].
\item 
  The voting areas that voted for K with a total of $7d_2$ votes, for
  any digit $d_2$, show an unusual characteristic: 15 of the 20 values
  are odd numbers, and the even numbers occur exactly once each
  (Table~\ref{t-Kseventies}). The chance of the former is 
  $\psevena \approx 0.04$ [Eq.~(\ref{e-psevena})]. However, a related
  coincidence is that the full set of K7-selected shahrestans is 
  dominated by odd vote counts for all four candidates, giving
  $\pKsevenallodd \approx 0.00044$ [Eq.~(\ref{e-k7-allodd})].
\end{list}

This sequence of significantly unusual properties of the MOI dataset 
is difficult to explain other than by artificial intervention in the data.
While it is true that all of these tests (except for the plan to use
a Benford's-Law-like analysis) are {\em post-hoc} 
tests, i.e. they were chosen after seeing
the data, a \v{S}id\`ak-Bonferonni correction \citep{Abdi07}
cannot help much. Combining the initial
local-bootstrap first-digit test, (1) the six biggest cities parametric test,
and (3) the odd dominance of the K7-selected shahrestan vote counts,
and a \v{S}id\`ak-Bonferonni correction of $C$ gives 
\begin{equation}
  \pall < C 
  % 1.5e-3    [Eq.~(\ref{e-pk7-sidakbon})].
  \pseven \;
  % 5e-4   [Eq.~(\ref{e-psevenabcGauss})].
  \psevenabcG \;
  % 0.00044 * 0.00037
  \pKsevenallodd
  % 1.5e-3 * 5e-4 * 0.00044 
= 3 \times 10^{-10} C. 
\end{equation}
For $\pall$ to be rejected with less than 95\% confidence, 
$C$ would have to be greater than $10^{8}$. It is difficult
to imagine that the number of similar statistical tests to those
used above is this large. A stronger objection to this estimate of 
$\pall$ is that these different tests may not be statistically
independent. However, it is difficult to imagine any natural
dependence between these three tests, which each give $p \sim 10^{-3}$ 
or lower {\em individually}, that could bring the probability of their combined 
occurrence to anywhere near unity.

While the first-digit analysis and subsequent unusual characteristics 
of the MOI data are difficult to explain other than as an effect of
artificial interference, these (mostly) do not lead to any indications
of what the unaltered data would look like, apart 
from the six-big-cities K7 selection, in which K7-selected vote 
counts for A can be interpreted as a misestimate
of the true vote, suggesting that
that the difference between A's and M's vote totals is overestimated
by about one million votes.

The set of pre-election opinion polls presented and analysed above 
in Table~\ref{t-prepolls}, Fig.~\ref{f-prepolls}, and
Tables~\ref{t-prepolls-prob} and \ref{t-prepolls-best} 
gives qualitatively similar results. This list of polls appears
to be that which was constructed by what is likely to be the least
biased (not necessarily unbiased) method available to the 
English-speaking community. The Wikipedia article in which the list
was originally compiled included 345
edits to the English-language Wikipedia article up to 11
June 2009 (inclusive) \citep{Gerash772009b}, 969 edits to the
corresponding Persian-language Wikipedia article \citep{Agha2009} up to
the same date, each by about 84 individual human authors, seven of
whom edited both articles, and the number of readers who found the
information sufficiently accurate and reflecting a neutral point of
view that they declined to edit the articles
is certainly much larger \citep{Gerash772009a,Gerash772009b}. 
Linear least-squares fits to the poll set show the following.
%(Sections~\ref{s-prepolls-data} and \ref{s-prepolls-res}) 
%\begin{list}{(\roman{enumi})}{\usecounter{enumi}}
\begin{list}{(\arabic{enumi})}{\usecounter{enumi}}
\item
  No matter which subset of polls is chosen, the linear least-squares
  fits to the polls internally agree best on the vote intention
  evolution either for candidate K alone or for the implied RKO (R
  plus K plus ``other'') vote intentions compared to either of
  the major candidates, A and M (or R alone in the former case)
  ($p_{\chi^2}$, Table~\ref{t-prepolls-prob}).
\item
  The official results are rejected significantly or to very
  high significance for nearly every candidate (or RKO) in 
  nearly every subset of polls 
  ($p_{y_j}$ in Table~\ref{t-prepolls-prob}, and 
  Table~\ref{t-prepolls-best}). 
\item
  Even the most conservatively selected subset of polls, that
  excludes partially internet-based and publicly unarchived
  polls and selects those remaining polls that give an
  absolute majority to candidate A (fifth row of the two tables),
  rejects the official combined vote for candidates R and K 
  with $p \approx 10^{-4}$, provided that most or all of the
  polls are normalised (undecided/other responses excluded), which
  appears to be the case from the available data.
\item
  The poll subset favouring A (fifth row of the two tables) implies a
  highly significant win for A, while the subset disfavouring A (sixth
  row of the two tables) implies a weakly significant win for M. 
\item 
  Unless the subset favouring or disfavouring A is chosen, the best
  estimate of the election result is that both A and M received less
  than or just slightly over 50\% of the votes ($\sim
  35$--$40\;\pm 7\%$ and $\sim 40$--$50\;\pm 5\%$, respectively).
  Neither a first-round win by A, a first-round win by M, or a need
  for a second round election are significantly excluded.
\item
  Unless the subset favouring or disfavouring A is chosen, the most
  likely correct result was that neither candidate won the 
  election first round.
\end{list}

Several possible interpretations of (1)--(3)
are listed in Section~\ref{s-prepolls-res-RKO}. If the partially 
internet-based polls overestimated K's support by about a 
factor of 10 and if most of the polls that do not state support levels
for K are not normalised, then the polls would be consistent
with the official results. However, those polls that state support
for all the candidates are normalised.

If there was a swing of about 90\% against K or about 70--90\% against
R and/or K during the last several days of the election, i.e. if {\em
  nearly all} of their supporters suddenly changed their minds just before the
election, then this also would be consistent with the official
results.  However, the data that are available for K show support consistent
with being constant over three months, suggesting that a 70--90\% loss
of support would require an extremely sudden, strong discouragement of
previously loyal supporters.

The most consistent way to explain these results would appear to be
the hypothesis of artificial interference in the official results.  In
this case, if the best estimates for the correct results are obtained
from the pre-election polls alone, then this would give about 7\%
support to candidate K or about 10--20\% support to candidates R and K
together in the 12 June first-round results
(Table~\ref{t-prepolls-best}), in contrast with the official result of
2.5\% for the combined vote for R and K.  This would be consistent
with the first-digit analysis and subsequent unusual properties of the
official results, which point directly to extremely unusual properties
starting from the excess of vote counts for K that start with the
digit 7.

Further evidence for or against either the hypothesis of no interference
in the data or the opposite could potentially be obtained by
examining the vote counts in the six largest shahrestans
(Table~\ref{t-Kseven}), and in odd-dominated K7-selected and 
shahrestans 
(Table~\ref{t-K7-all}).
The voting areas' names are listed in the table published by the MOI
\citep{MOI09,MOI09b,MOI09c}. 

Although the local bootstrap method presented here
has led to the detection of highly 
significant anomalies in an electoral poll,
{\em the reverse would not be true in general.} Benford's-Law-like
tests may
detect anomalies in a dataset but cannot guarantee the absence of
anomalies, because many randomising effects can combine to hide
artefacts. The anomalies detected in the analysis presented above,
where the big city effect suggests an error of about one million votes, 
may not constitute
the full set of anomalies, nor do they provide an estimate of 
the unaltered data.
The compilation of pre-election poll data, despite 
relatively poor reporting quality in individual sources available directly
or indirectly in the English language, probably together provide the most
robust estimates available of the unaltered data. Higher quality 
poll reporting, in particular including data for all candidates
and undecided/other percentages, as well as detailed methodologies, 
would reduce the ambiguities remaining in analyses of pre-election data.

Regarding the Iranian 2009 presidential-election first round in 
particular, at least three
statistical analyses independent of this work have been carried out
\citep{Meb09a,BebScacco,Chatham09}.
Based on very different methods of analysis and to some degree
on different data (e.g. including the 2005 election results and
excluding pre-election opinion polls), 
the analyses of Mebane \citep{Meb09a} and Chatham House \citep{Chatham09}
come to conclusions that are compatible with the present results,
in particular regarding the unusual nature of the official results
for candidate K.

\section*{Acknowledgements}
Thank you to the pseudo-anonymous Wikipedia editor ``128.100.5.143'' 
who alerted me to the MOI publication of the data set,
to Walter R. Mebane Jr for providing a copy of the 2005 Iranian data set,
to an anonymous physicist from Iran for a useful idea,
and to Maziar Parizi for providing a copy of his 
 private circulation preprint, 
``Le miracle de l'\'election Iranienne de juin 2009'' (v05\_01).
Thank you to numerous people who provided useful comments and will remain
anonymous unless they request otherwise. 
\postrefereechanges{Thank you to the referees for useful comments,
which have led to several improvements in the
paper.}
Thank you to Stephen Stigler for suggesting the comparison with similar
empirical data.
This work has used the GNU {\sc Octave} command-line, high-level
numerical computation software (\url{http://www.gnu.org/software/octave}).

%MAINTEXT word count

%\bibliographystyle{acmtrans-ims}
%\bibliography{mybib}

\tabfiglater{
%%% TABLES %%%
  \clearpage
  \tpreselections	\clearpage
  \tprepolls	\clearpage
  \tprepollsprob	\clearpage
  \tprepollsbest	\clearpage
  \tprepollsbestlinear  \clearpage
  \tprepollsbestquadratic  \clearpage
  \tprepollsbestcubic  \clearpage

  \tKseven	\clearpage
  \tKseventies	\clearpage
  \tKsevenall	\clearpage
  \tpollobservers	\clearpage

%%% FIGURES %%%
  \fsigabs	\clearpage
  \fsigrel	\clearpage
  \fhistall	\clearpage

  \fhistA	\clearpage
  \fhistR	\clearpage
  \fhistK	\clearpage
  \fhistM	\clearpage

  \fcalib	\clearpage
  \fbooteg	\clearpage

  \fAconflev	\clearpage
  \fRconflev	\clearpage
  \fKconflev	\clearpage
  \fMconflev	\clearpage

  \fprepolls  	\clearpage %% two panels = two files
  \fhistKfolded	\clearpage
  \fKsevenA	\clearpage
  \fKsevenR	\clearpage
  \fKsevenK	\clearpage
  \fKsevenM

}

\end{document}